\documentclass[sigconf,nonacm,balance=false]{acmart}
\usepackage{color, colortbl}
\usepackage{array}
\usepackage{booktabs,caption}
\usepackage[flushleft]{threeparttable}
\usepackage{makecell}
\usepackage{wrapfig,lipsum,booktabs}
\usepackage{array}
\newcolumntype{M}[1]{>{\centering\arraybackslash}m{#1}}
\usepackage{pstricks-add}%
\usepackage{subcaption}
\usepackage{framed}
\usepackage{varwidth}
\usepackage{graphicx}
\usepackage{grffile}
\usepackage{multicol}

\newcounter{rcounter}
\newenvironment{rec}[1][]{\refstepcounter{rcounter}\par\itshape\medskip
   \noindent\textit{\textbf{Recommendation \thercounter:} #1}}{\medskip}

\begin{document}

\title[Data Safety vs. App Privacy]{Data Safety vs. App Privacy: Comparing the Usability of Android and iOS Privacy Labels}


\author{Yanzi Lin}
\affiliation{%
  \institution{Wellesley College}
  \city{Wellesley}
  \state{Massachusetts}
  \country{USA}}
\email{yl102@wellesley.edu}

\author{Jaideep Juneja}
\affiliation{%
  \institution{Carnegie Mellon University}
  \city{Pittsburgh}
  \state{Pennsylvania}
  \country{USA}}
\email{jjuneja2@alumni.cmu.edu}

\author{Eleanor Birrell}
\affiliation{%
  \institution{Pomona College}
  \city{Claremont}
  \state{California}
  \country{USA}
}
\email{eleanor.birrell@pomona.edu}

\author{Lorrie Faith Cranor}
\affiliation{%
 \institution{Carnegie Mellon University}
 \city{Pittsburgh}
 \state{Pennsylvania}
 \country{USA}}
\email{lorrie@cmu.edu}

\renewcommand{\shortauthors}{Lin et al.}

\begin{abstract}
Privacy labels---standardized, compact representations of data collection and data use practices---are often presented as a solution to the shortcomings of privacy policies. Apple introduced mandatory privacy labels for apps in its App Store in December 2020; Google introduced mandatory labels for Android apps in July 2022. iOS app privacy labels have been evaluated and critiqued in prior work. In this work, we evaluated Android Data Safety Labels and  explored how differences between the two label designs impact user comprehension and label utility. We conducted a between-subjects, semi-structured interview study with 12 Android users and 12 iOS users. While some users found Android Data Safety Labels informative and helpful, other users found them too vague. Compared to iOS App Privacy Labels, Android users found the distinction between data collection groups more intuitive and found explicit inclusion of omitted data collection groups more salient. However, some users expressed skepticism regarding elided information about collected data type categories. Most users missed critical information due to not expanding the accordion interface, and they were surprised by collection practices excluded from Android's definitions. Our findings also revealed that Android users generally appreciated information about security practices included in the labels, and iOS users wanted that information added. 
\end{abstract}


\maketitle
\section{Introduction}
\label{sec:Introduction}

Websites and mobile apps provide prospective users with information about their data use practices---typically through a privacy policy---and people who subsequently choose to interact with a website or download an app are presumed to have consented to the described data practices. However, extensive research has shown that privacy policies are an insufficient basis for informed consent. Privacy policies are long, employ complex terminology, and contain vague and inconsistent language~\cite{chen2021fighting,mcdonald2008cost,mcdonald2009comparative,vila2004we,reidenberg2016ambiguity,singh2011evaluating,korunovska2020challenges,zhangprivacy}; as a result, many individuals struggle to accurately understand the described data practices. 
%
\emph{Privacy labels}---compact representations that display key data practices in a standardized, comprehensible format---are one approach for ameliorating the shortcomings of privacy policies. Prior work has designed and evaluated privacy labels for websites~\cite{kelley2009nutrition,kelley2010standardizing}, mobile apps~\cite{kelley2013privacy}, and IoT devices~\cite{emami2019exploring,emami2020ask,emami2021informative}. These studies suggest that privacy labels might more effectively inform users about data use practices compared to privacy policies, thereby constituting a better basis for empowering informed consent. 

Apple introduced mandatory ``App Privacy'' labels for apps available through its App Store in December 2020~\cite{inc._2020}. However, user studies evaluating  these iOS app privacy labels found that many people misunderstand them~\cite{zhang2022usable,kollnig2022goodbye} and that user satisfaction with the label design is low~\cite{zhang2022usable}. Prior work has also established that developers misinterpret terms and make common errors when creating labels~\cite{li2022understandingB}, resulting in many inaccurate labels~\cite{koch2022keeping,xiao2022lalaine,jain2023atlas}. Google introduced its own mandatory privacy labels---which it calls ``Data Safety'' labels---with labels required for all apps by July 2022~\cite{googleLabelBlog}. While prior work has found that many apps have inconsistencies between their iOS and Android labels~\cite{khandelwal2023overview}, the usability of Android data safety labels has not previously been studied in the academic literature. 

This paper extends prior work by evaluating the usability of Android Data Safety Labels and comparing Google's label design with the iOS label design, focusing on how design differences between Android and iOS labels affect user comprehension and perceptions of the labels. 

We conducted a between-subjects, semi-structured exploratory interview study with 24 participants: 12 Android users and 12 iOS users. We asked background questions and then showed each participant a series of three privacy labels; Android users saw the Android Data Safety Label and iOS users saw the iOS App Privacy Label for each app. For each label, we observed how participants interacted with the label, we asked a series of factual questions to evaluate user comprehension, and we asked questions about users' impressions of the app. We concluded with questions about overall impressions about privacy labels and mobile app privacy. 

\begin{figure*}
\begin{subfigure}{.49\textwidth}
\begin{framed}
    \centering
    \includegraphics[width=\textwidth]{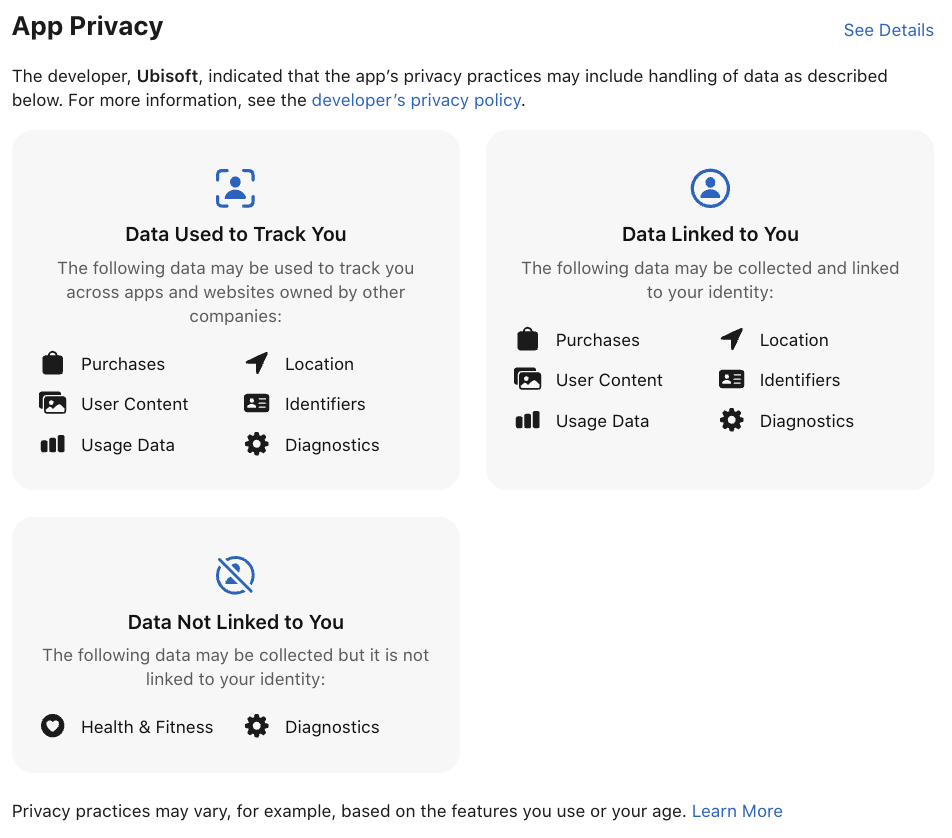} 
\end{framed}
\caption{iOS App Privacy Label}\label{subfig:compact_ios}
\end{subfigure}
\begin{subfigure}{.49\textwidth}
\begin{framed}
    \centering
    \includegraphics[width=\textwidth,height=1.0056cm]{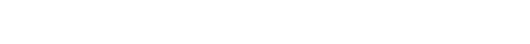}
    \includegraphics[width=\textwidth]{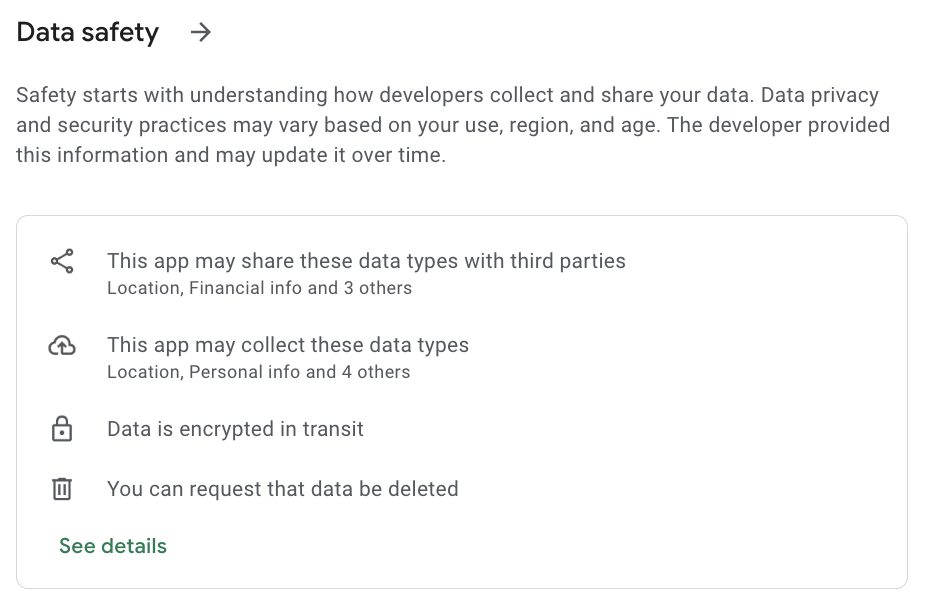} 
    \includegraphics[width=\textwidth]{figures/whitespace.png}
\end{framed}
    \caption{Android Data Safety Label}\label{subfig:compact_android}
\end{subfigure}
\caption{Compact labels for the Just Dance Now mobile app.}\label{fig:compact_labels}
\end{figure*}

We found that Android Data Safety Labels suffered from many of the same or similar problems as have previously been identified for iOS labels, including problems with misunderstanding terms and uncertainty about whether the labels can be trusted. While some users find Android labels informative and helpful, other users find them too vague. 

We also identified differences. Compared to iOS App Privacy Labels, Android users found the distinction between data collection groups more intuitive and found explicit inclusion of omitted data collection groups more salient. However, some users were suspicious of elided information about which data type categories were collected, most users missed critical information because they didn't expand the accordion interface, and most users were surprised by collection practices excluded from Android's definitions. We also found that Android users generally appreciated the information about security practices included in the app privacy label, and iOS users wanted that information added. 

Based on our results, we identify concrete recommendations for design improvements that could be adopted by mobile app stores, and we discuss areas for future research.

\section{Background}\label{sec:background}

Apple and Google introduced privacy labels for mobile apps in December 2020 and April 2022, respectively. Developers who release a new app or who update an existing app must provide information for the label; the app store then shows a privacy label on the download page for each app. Both companies have promoted their labels as an effort to enhance transparency about the data practices of mobile apps, and the two label designs have some common elements. However, the two designs also exhibit unique elements and design features. This section briefly summarizes and compares the key design elements of iOS app privacy labels and Android data safety labels as of the time of our study in August 2022. In the following year, we did not observe any changes to the label designs.

\subsubsection*{iOS App Privacy Labels.} iOS privacy labels, released in 2020~\cite{cylab_article_ios},  have two forms: the compact label shown on the download page of each mobile app and the expanded label accessible through a blue ``See Details'' link in the top right corner of the compact label. 

The compact iOS App Privacy Label categorizes data collection into three data collection groups: ``Data Used to Track You,'' ``Data Linked to You,'' and ``Data Not Linked to You.'' Tracking is narrowly defined as data used for targeted advertising or shared with a data broker. Data linked to you includes any information tied to a user's identity and explicitly includes any personal information as defined by ``relevant privacy laws.'' Data not linked to you must not be associated with any identifier linked to a real-world identity and it must not be associated with other data or used in a way that would allow users to be re-identified. For each of these three groups, the compact label shows the full list of data type categories in that group; if no data type categories fall into a group, that group is omitted from the compact label. The compact label also includes a blue link to the app developer's privacy policy near the top and a blue ``Learn More'' link to information about app privacy at the bottom. An example compact iOS label is shown in Figure~\ref{subfig:compact_ios}.

\begin{figure*}
\begin{subfigure}{.4\textwidth}
    \begin{framed}
        \centering
        \includegraphics[width=\columnwidth]{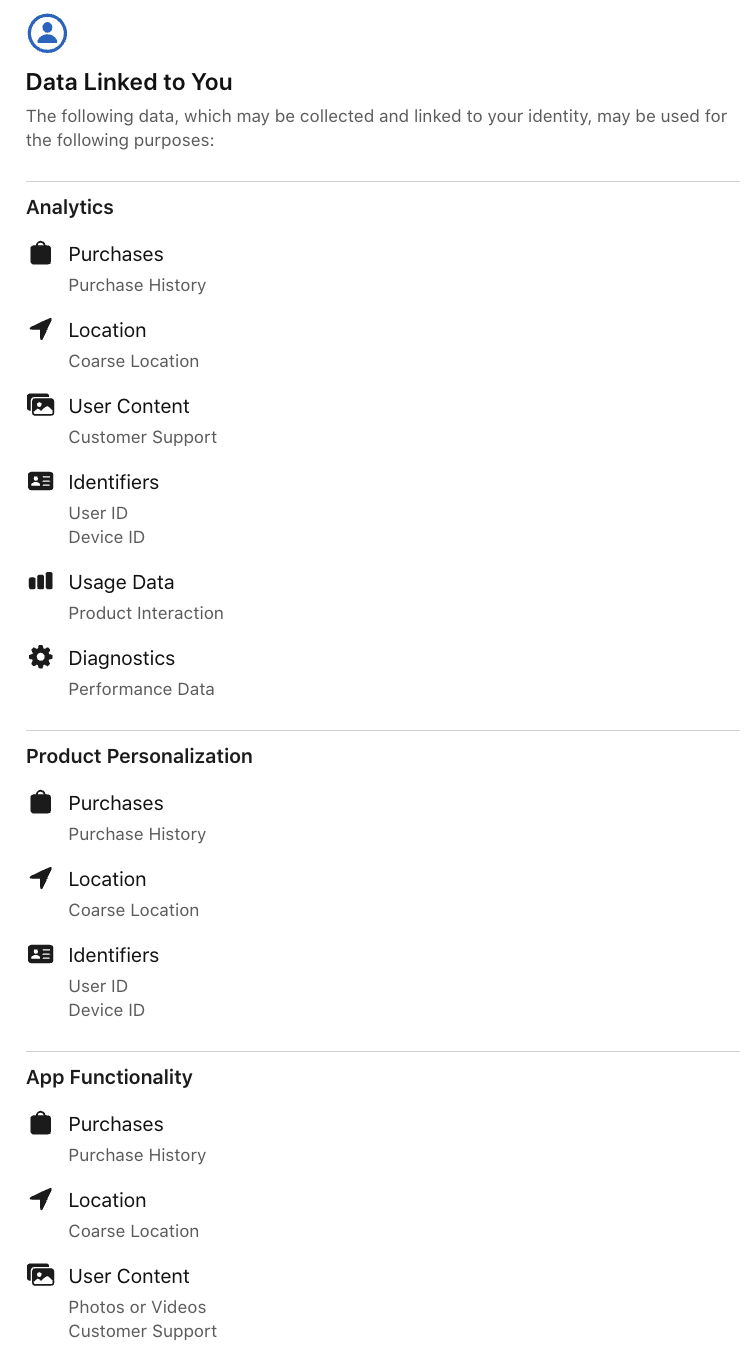}
    \end{framed}
    \caption{iOS App Privacy Label}\label{fig:expanded_ios}
\end{subfigure}
\begin{subfigure}{.4\textwidth}
    \begin{framed}
        \centering
        \includegraphics[width=\columnwidth]{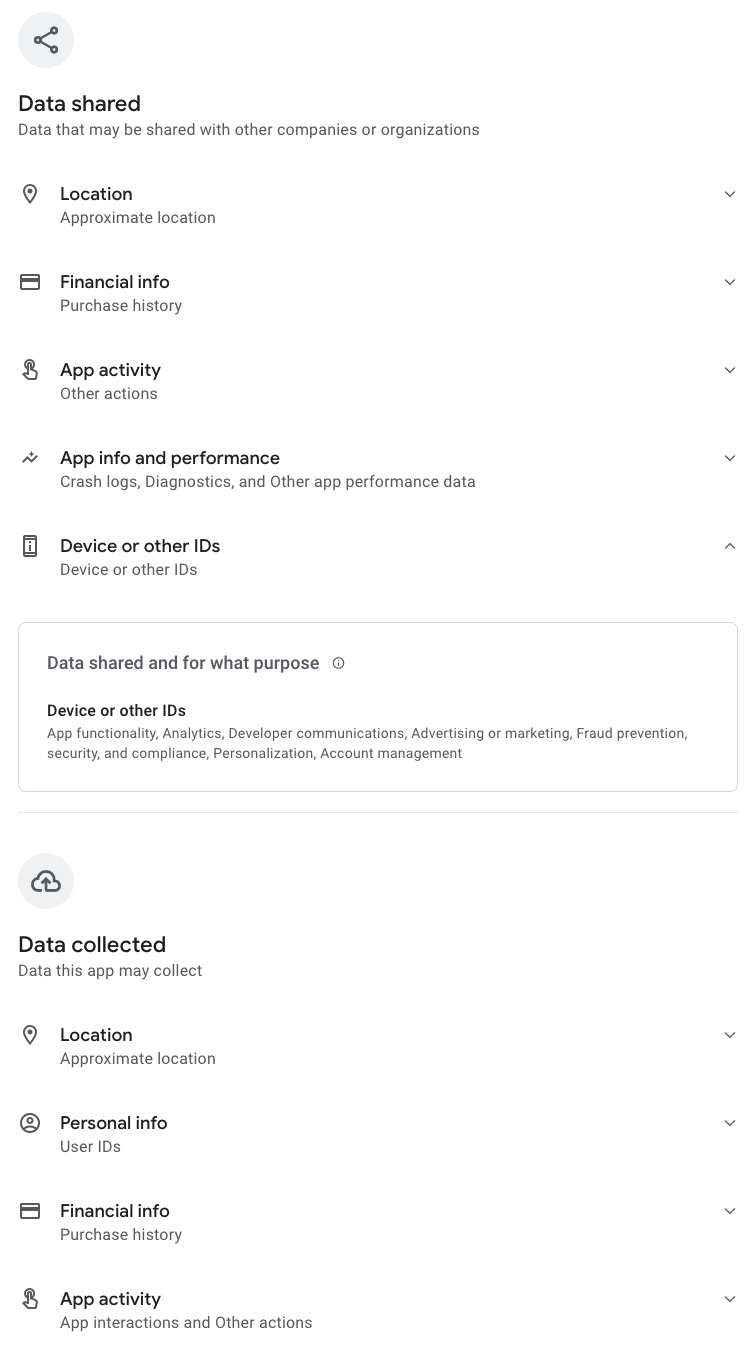}
    \end{framed}
\caption{Android Data Safety Label}\label{fig:expanded_android}
\end{subfigure}
\caption{Excerpts from the expanded labels for the Just Dance Now mobile app.}
\end{figure*}

The expanded iOS app privacy label provides additional information about data practices. In the expanded label, specific data types are listed under each data type category (e.g., ``Photos or Videos'' and ``Customer support'' under the data type category ``User Content'' in Figure~\ref{fig:expanded_ios}).  Within the ``Data Linked to You'' and ``Data Not Linked to You'' groups, the expanded label lists data type categories for each group under one or more purposes. For example, ``Data Linked to You'' might include three data uses---analytics, product personalization, and app functionality---with a subset of the data type categories linked to you listed under each usage. If a data type category is used for multiple purposes, it will be listed under each of the applicable purposes (possibly with different specific data types if those types are not used for all purposes). The expanded label also includes a blue link labeled ``Privacy Definitions and Examples'' that links to definitions and explanations of key terms used in the iOS label. An excerpt from an example expanded iOS label depicting the ``Data Linked to You'' section of the expanded label is shown in Figure~\ref{fig:expanded_ios}. Full labels are included in Appendix~\ref{appendix:label_screenshots}.


\subsubsection*{Android Data Safety Labels.} Android Data Safety Labels, released in 2022~\cite{wp_article_android}, also have a compact label and an expanded label, with the expanded label accessible through a green ``See details'' link at the bottom of the compact label.

The compact Android Data Safety Label categorizes data collection into two groups that do not align with the iOS groups: data shared with third parties and data collected. Data transferred to a third party is generally considered ``shared,'' but there are exceptions for data shared with service providers, data shared after explicit user consent, and anonymized data. Data collection is defined as data that is retrieved off the local device, although there are exceptions for ``ephemeral'' processing and for data sent using end-to-end encryption. 
Unlike the iOS compact label that shows the full list of data type categories for each group, the Android compact label lists only the first two data type categories and the number of additional data type categories (unless the total number of data type categories for that group is three or fewer, in which case the compact label lists all of them). If there are no data type categories in a group, the compact label explicitly states so, unlike the iOS label, which omits groups that have no data type categories.

The compact label also specifies whether or not the app encrypts data in transit and whether or not a user can request to have their data deleted. If an app has committed to follow the ``Play Families Policy'' or has undergone an independent security review, that information is also displayed in the compact label. An example compact Android data safety label is shown in Figure~\ref{subfig:compact_android}.

In the expanded label, specific data types are listed under each data type category (e.g., ``App interactions'' and ``Other actions'' under the data type category ``App activity''). 
While the expanded iOS label lists data practices with specific data types below each practice, the expanded Android label lists specific data types in an accordion interface that can be expanded to show data practices. 
The expanded label also includes a link to the app developer's privacy policy and a link labeled ``Learn More'' that links to definitions and explanations of key terms used in the Android label. An excerpt from an example expanded Android label---depicting the ``Data shared'' group of the expanded label with one data type category (``Device or other IDs'') clicked---is shown in Figure~\ref{fig:expanded_android}. Full labels are included in Appendix~\ref{appendix:label_screenshots}.


\subsubsection*{Key Design Differences} Both iOS App Privacy Labels and Android Data Safety Labels aim to enhance transparency about the data use practices of mobile apps. However, the look and feel of their labels is different and they use differing terminology. For example, Apple employs the term ``Data Use'' while Google uses ``Data Purposes'' to mean the same thing. In this paper, we will adhere to these respective terminologies and refer to either platform using the appropriate terminology. However, when discussing both privacy labels collectively, we will use the term ``data collection purposes.'' Here we highlight the key differences between the design of the two privacy labels that we will revisit as we analyze our data:

    \textbf{Location and formatting of links.} ``See Details'' appears in blue on the top right corner of the iOS compact label. It appears in green at the bottom of the Android compact label.
    In iOS, the link to the developer's privacy policy appears prominently at the top of the iOS labels (both compact and expanded) and appears only at the very bottom of Android expanded labels. 
    In iOS a link to ``Privacy Definitions and Examples'' appears in blue near the top of the expanded label. Android provides a gray ``Learn More'' link near the top of the expanded label that leads to definitions.
    iOS also has a blue ``Learn More'' link near the top of the expanded label that leads to information about app privacy, but not definitions.

    \textbf{Data collection groups.} iOS labels have three data collection groups: ``Data Used to Track You,'' ``Data Linked to You,'' and ``Data Not Linked to You.'' In contrast, Android labels have two data collection groups: ``Data Shared'' and ``Data Collected.'' These two approaches to categorizing data address somewhat different characteristics of data. The two labels also define different exclusions---i.e., types of collection that do not need to be reported in the privacy label---which impacts which data type categories are reported under each data collection group.  

    \textbf{Completeness of compact label groups and category lists.} The presentation of empty collection groups is handled differently in Android and iOS labels. iOS labels omit empty collection groups, whereas Android labels explicitly state that no data is collected in that group. 
    iOS labels provide an exhaustive list of all the data type categories in the compact label with icons displayed next to them.  Android labels provide a shorter and more compact look; if there are four or more data type categories for a data collection group,  the Android label displays only the first two and ``+$n$ others'' .  
    
    \textbf{Expanded label structure.} The expanded iOS label lists data uses, with the data type categories and specific data types used for that use displayed underneath each use. In contrast, the expanded Android label displays data type categories, with purposes and the list of specific data types used for that purpose displayed underneath each data type category when the accordion interface is unfolded. Thus the iOS label presents a purpose-centric view while the Android interface presents a data-centric view. In addition, the iOS label requires a lot of scrolling to navigate the expanded view, while the Android label is more compact but requires multiple clicks to expand the accordion interface.   

    \textbf{Representation of security and data deletion practices.} Both the compact and expanded Android labels display information about whether an app encrypts data in transit and whether it is possible to request to delete data. If an app complies with the Play Families Policy or if an independent security review has been conducted, that information is also displayed. There is no analogous information provided in iOS labels. 

\section{Related Work}
\label{sec:Related}

We present related work evaluating iOS and Android privacy labels, including an interview study similar to ours that evaluated only iOS labels~\cite{zhang2022usable}. In addition, we discuss related work evaluating other privacy-related communications, including privacy policies and privacy labels in other contexts.

\subsection{Evaluations of iOS and Android Labels}

Zhang et al. conducted an interview study with 24 iPhone users to explore people's understanding  and perceptions of iOS App Privacy Labels. They found that most users were unaware of iOS privacy labels. After looking at example labels, participants found them useful, although some considered them vague and many did not trust their accuracy. This work identified common misunderstandings of terms used in iOS privacy labels, including confusion about the definitions of data collection groups (e.g., ``Data Used to Track You'' and ``Data Not Linked to You'') and various data type categories (e.g., ``user content'' and ``identifiers'')~\cite{zhang2022usable}. Our work replicates many of the results found by Zhang et al. and extends their work by evaluating Android Data Safety Labels and comparing the two label designs.

Prior work has also evaluated iOS App Privacy Labels from the developer's perspective. Li et al. conducted an interview study with 12 iOS app developers to understand their challenges in creating labels. They identified common developer errors that led to inaccurate labels, including misunderstanding ``Data Linked to You,'' underestimating data use by third-party libraries, forgetting about collected data, incorrectly handling optional data practices, and incorrectly reporting locally-stored data. They recommended revised definitions, expanded documentation, and automated validity-checking tools to help developers create accurate privacy labels~\cite{li2022understandingB}. 

Several research teams have performed quantitative analyses of large datasets of iOS App Privacy Labels.
Li et al. analyzed a longitudinal dataset of 1.4 million apps on the U.S. App Store from April to November, 2021. They found that developers were slow to add privacy labels, with most apps only adding or updating their privacy label when they update the app. They also reported statistics about reported data practices~\cite{li2022understandingA}. 
Kollnig et al. performed a comparative analysis of 1,759 iOS apps before and after Apple introduced app privacy labels. They found that app privacy labels are inconsistent with actual data practices. For example, 80.2\% of apps that claimed they did not use data to track the user in fact contained ad libraries. They also concluded that the addition of app privacy labels had not changed apps' data use practices~\cite{kollnig2022goodbye}. 
Koch et al. performed a statistical analysis of 11,074 iOS apps and their privacy labels (or lack thereof); they found that most apps report collecting some data, and that game apps in particular collect more data and use more data for tracking. Many of the labels in their dataset include inconsistencies, for example, 13\% erroneously claim personal information as ``Data Not Linked to You''~\cite{koch2022keeping}.

Prior work has also analyzed applications to evaluate the accuracy of iOS App Privacy Labels. Koch et al. dynamically analyzed the traffic of 1,687 iOS apps and found that 16\% of apps transmit data without declaration~\cite{koch2022keeping}. Xiao et al. used a combination of static and dynamic analysis to systematically evaluate consistency between 5,102 iOS labels and actual app behavior; they found that 67\% of iOS labels are inconsistent with actual app behavior, with most apps failing to fully disclose all data collection and purposes~\cite{xiao2022lalaine}. Jain et al. used NLP techniques to compare iOS App Privacy Labels to app privacy policies; they analyzed  354,725 iOS apps and found that only 29.6\% of apps provided both an App Privacy label and a privacy policy and that 88.0\% of those apps exhibited possible discrepancies between the label and the privacy policy~\cite{jain2023atlas}.

The first work to look at Android Data Safety labels was a recent study by the Mozilla Foundation that looked at 40 top Android apps and found that almost 80\% had Data Safety Labels that were inconsistent with the app's privacy policy~\cite{mozilla_2023}. Khandelwal et al. conducted a large-scale analysis of 165K apps listed on both iOS and Android app stores; they found that many apps were missing labels and that most apps with labels had inconsistencies between the two labels~\cite{khandelwal2023overview}. In other work, Khandelwal et al. conducted a large-scale longitudinal analysis of 1.14 million Android apps and found that missing labels and internally inconsistent labels are common, and that many apps are updating and refining their labels over time. They also identified challenges based on survey responses from 889 Android developers~\cite{khandelwal2023unpacking}.

\subsection{Evaluations of Privacy Communications}
Although mobile app privacy labels have only existed for a few years, the idea of privacy labels has been around for over two decades, motivated by concerns about the length of privacy policies and their ability to communicate to users~\cite{mcdonald2008cost}. As early as 2001, Commissioners of the U.S. Federal Trade Commission advocated for standardized privacy ``nutrition labels'' to help consumers understand and compare website privacy practices~\cite{anthony2001case}. Since then, privacy labels have been proposed and evaluated in a variety of domains including websites, mobile apps, and IoT devices. 

Kelley et al. first proposed designs for ``privacy nutrition labels'' in 2009 in the context of website privacy policies. They designed and evaluated  standardized labels through a series of focus groups, lab studies, and online surveys. Their final design---which centered a matrix depicting practices for data type-purpose pairs---allowed users to more quickly and accurately identify a website's data practices compared to natural-language privacy policies~\cite{kelley2009nutrition,kelley2010standardizing}. 

In 2009, eight US federal agencies collaborated on a model privacy notice for US financial institutions that included a table of bank data practices in a standard format, similar to a nutrition label. Cranor et al. collected over 6000 bank privacy notices in this format and conducted a large-scale analysis. They also identified issues with the model notice design~\cite{Cranor2016bank}.

Later work extended privacy labels to other domains. Kelley et al. proposed ``Privacy Facts'' for mobile apps that swayed users towards selecting more private apps in contexts where the user was choosing between otherwise similar apps~\cite{kelley2013privacy}. More recently, Emami-Naeini et al. proposed a privacy label for IoT devices; they found that their proposed privacy label effectively communicated information about privacy risks to users and that it influence hypothetical user purchasing decisions~\cite{emami2019exploring,emami2020ask,emami2021informative}. 

Cookie banners are another type of privacy-related communication that has been evaluated by researchers. Researchers have found that consumers largely do not understand the privacy decisions facilitated by these banners and that the banners often nudge users to make choices that do not align with their preferences~\cite{Habib2022cookie,Bouma-Sims2023cookie}. 

Habib and Cranor synthesized approaches used to evaluate privacy choice interfaces and proposed an evaluation framework with seven factors~\cite{Habib2022evaluating}. Although app privacy labels do not directly offer choice interfaces, they may be used to inform users' app download decisions. Thus, these factors are relevant to our evaluation, especially two of the factors: awareness and comprehension.

\section{Methods}
\label{sec:Methods}
We conducted a semi-structured interview study in which we presented participants with privacy labels for three mobile apps and asked them factual questions about each app's data use practices along with questions about the participant's impressions and opinions about the label design. We recruited both Android and iOS users; Android users were shown the Android privacy labels for the three apps and iOS users were shown the iOS privacy labels for the three apps. 
We elected for a between-subjects design to avoid priming and learning effects, as well as to allow participants to interact with labels on their own devices in the style used by their phone's OS. A within-subjects design that exposed participants to two different label designs for the same app would risk priming participants to form expectations for the second label based on what they saw in the first label. In addition, participants might learn where to find information on the first label, and thus either have an easier time finding it on the second label if it is in a similar place, or have a harder time finding it if it is in a different place. 
Additionally, participants are more likely to provide informed responses by interacting with privacy labels on their own devices due to their familiarity with the respective interfaces.



The Android and iOS interview scripts can be found in Appendix~\ref{sec:appendix_android_scripts} and Appendix~\ref{sec:appendix_ios_scripts}. We developed a first draft of our interview script and piloted it with three volunteers. We then made adjustments to clarify questions and reduce the interview duration. Our final interview protocol was designed to take approximately 1 hour. Actual interviews ranged from 51 to 77 minutes and on average lasted 63 minutes. 

\subsection{Privacy Label Selection}
We examined privacy labels of popular free apps in the Google Play Store and the Apple App Store to identify apps whose labels encompassed a diverse range of features, providing a comprehensive overview of each app's privacy practices. Specifically, we wanted to include at least one app that exhibited each of the following features:
\begin{enumerate}
    \item \textbf{All iOS collection groups:} The app's iOS label reports at least one data category under each of the three iOS data collection groups (``Data Used to Track You'', ``Data Linked to You'', and ``Data Not Linked to You'').
    \item \textbf{Missing iOS collection group:} The app's iOS label has at least one group under which no data categories are collected.
    \item \textbf{All Android collection groups:} The app's Android label reports at least one data category under each of the two Android data collection groups (``Data Shared'' and ``Data Collected'').
    \item \textbf{Missing Android collection group:} The app's Android label has at least one group under which no data categories are collected.
    \item \textbf{All Android security features:} The app's Android label states that data is encrypted in transit and that users can request that data be deleted.
    \item \textbf{Missing Android security feature:} The app's Android label either states that data is not encrypted in transit or that users cannot request that data be deleted. 
    \item \textbf{Android security review:} The app's Android label indicates that the app underwent an independent security review. 
\end{enumerate}
To select our apps, we reviewed the list of apps that appeared on the top free app charts in both app stores and that provided both iOS and Android privacy labels. We ultimately selected three apps: (1) Just Dance Now, (2) Google Maps, and (3) Tumblr–Fandom, Art, Chaos. Together, these three apps exhibited all of the desired features (Table~\ref{tab:appLabelFeatures}).

\begin{table}
\caption{Summary of the privacy label features exhibited by the three apps used in our study: Just Dance Now (JDN), Google Maps, and Tumblr – Fandom, Art, Chaos (Tumblr). }\label{tab:appLabelFeatures}
\begin{tabular}{l|ccc}
\hline
                            & JDN           & Google Maps   & Tumblr        \\
\hline\hline
1. All iOS groups           & $\checkmark$  &               &               \\
2. Missing iOS groups       &               & $\checkmark$  & $\checkmark$  \\
3. All Android groups       & $\checkmark$  &               &               \\
4. Missing Android groups   &               & $\checkmark$  & $\checkmark$  \\
5. All Android security     & $\checkmark$  & $\checkmark$  &               \\
6. Missing Android security &               &               & $\checkmark$  \\
7. Android security review  &               & $\checkmark$  &               \\
\hline
\end{tabular}
\end{table}

Rather than randomizing the order in which we presented the app labels, we selected a fixed ordering that allowed us to first observe participants using an app with a label that included a large number of typical features, and then to observe participants using apps with labels exhibiting more atypical features. Randomizing the order would introduce inconsistent priming effects, causing users to form varying expectations about the subsequent labels based on what they see in the first label. We showed the Just Dance Now labels first since this app's labels exhibited the features typical of most apps: all data collection groups were present in both labels, both standard security features were present, and no independent review was conducted. We then followed with the Google Maps and Tumblr apps to explore how atypical features --- missing data collection groups, independent security review, and missing security features --- impacted users. 

\subsection{Participant Recruitment} 

We used the crowd worker platform Prolific to recruit a total of 24 participants for our study: 12 iOS users and 12 Android users. We recruited participants for our interviews until we reached saturation. Towards the end of our interviews, we were not seeing new themes emerge so we stopped recruiting participants.

We required participants to be at least 18 year-old, located in the United States, fluent in English, and own either an Android or iOS phone to be eligible for our study. They were asked to read and consent to our online consent form and answer questions related to the eligibility criteria. Those who met the criteria were asked to fill out a short demographic survey (see Appendix~\ref{sec:appendix_dem_survey}) and were then redirected to a Calendly page where they could reserve a time slot for the Zoom interview. We paid participants \$0.40 for filling out the Online Consent Form and scheduling an interview. 
After the completion of the interview,  participants were compensated \$14.60 after the initial 60-minute interview. For interviews that ran long, participants received an extra payment of \$3.75 for every 15 minutes beyond the elapsed 60-minute interview. 

We utilized a purposive sampling method~\cite{tongco2007purposive} to ensure a balanced sample in terms of gender and age. Initially, we made 20 spots available to potential participants through Prolific. We then interviewed those who successfully completed the consent form and scheduled an interview with us. Since some participants were unable to schedule an interview, we continued to offer more spots on Prolific. However, we adjusted our qualification criteria to target individuals that identified with specific gender and age groups that were not yet represented in our study population. This approach allowed us to ensure that only participants meeting our desired criteria were able to view and respond to our recruitment. While our sample is not representative, our purposive sampling approach allowed us to observe labels being used by diverse participants within the constraints of a small sample.

\subsection{Interview Protocol}

Each interview was conducted by one of two researchers in August 2022. Twenty-three participants conducted the interview on their phones, and one participant conducted the interview on their laptop due to a technical difficulty they encountered with screen sharing through Zoom. 

After providing each interviewee with instructions, we asked a series of questions about their background. We then directed them to the privacy label for the first app (Just Dance Now). We gave them time to explore the label, asked about general impressions, and then asked specific factual questions to explore how well they understood the app's data practices. We then proceeded to ask questions relating to our second app (Google Maps) and third app (Tumblr). Finally, we asked participants to share their overall impressions of the labels, discuss whether or not they would look at labels in the future, and provide any other feedback or suggestions. 


\paragraph{Initial Instructions.} At the beginning of the interview, we requested permission from participants to record their phone screen while keeping their video off on Zoom to avoid capturing their face during the interview. Additionally, we informed participants that they could stop the interview at any time or decline to answer a question. We then gave participants the opportunity to ask the researcher any questions before proceeding. We reminded participants not to disclose any personally-identifiable information during the interview.

\paragraph{Background Questions.} We asked participants about their app download behavior and the criteria they consider when deciding to download an app. To assess participants' familiarity with privacy labels, we asked whether they would visit their platform's app store 
to learn more about an app's data practices, without specifically mentioning ``privacy'' or ``privacy labels.'' 

\paragraph{General Usage and Impressions (Just Dance Now).} Before introducing participants to the first label, we asked about their knowledge of the functionality of the Just Dance Now app and whether they had used it in the past. For those unfamiliar with the app, we provided a brief explanation of its functionality. Next, we directed participants to the compact version of the first privacy label (Just Dance Now) and told them to explore the label at their own pace. We recorded whether participants clicked on ``See Details'' on the compact label to see the expanded label of Just Dance Now unprompted. 
We then asked whether they had seen a compact privacy label like this before;  for those who had, we asked whether they had seen the label expanded as well. For Android users who viewed the expanded label, we recorded the number of data types they clicked on to expand the accordion interface, revealing the data collection purposes. 

We  asked participants questions about their overall impression of the information they saw and whether they found any information surprising or unexpected. 

\paragraph{Definition Questions (Just Dance Now).} We asked participants to explain the meaning of each data category and other key terms. Android users were asked about ``data sharing,'' ``data collection,'' ``encryption,'' and ``deletion.'' iOS users were asked about ``Data Used to Track You,'' ``Data Linked to You,'' and ``Data Not Linked to You.'' 

\paragraph{Impressions of the Expanded label and Comprehension (Just Dance Now).} We directed users to open the expanded label for Just Dance Now and told them to take some time to look it over. We then asked about impressions of the expanded label and anything that surprised them. We next asked a series of factual questions to test comprehension of the app's data practices. 

\paragraph{Omitted Collection Groups and Security Review (Google Maps).} We first assessed participants' familiarity and usage of the Google Maps app. Next, we showed participants the privacy label of Google Maps and asked them to compare it to the label of Just Dance Now and explain the differences. In comparison to Just Dance Now, the Google Maps label only includes the category of ``Data Linked to You'' on its iOS label. However, under this category, Google Maps collects a much larger range of data types. On its Android privacy label, Google Maps did not share any data, but it collected more data types. The Android label also shows that Google Maps passed an ``Independent Security Review.'' After the participant discussed initial impressions, we asked directed follow-up questions focusing on these differences. 

\paragraph{Security Practices (Tumblr).} Again, we began by asking about the participants' understanding of the app's functions and their usage history. Then, we introduced the privacy label of Tumblr – Fandom, Art, Chaos and asked participants to compare it to the two previous labels. In this section, our primary focus was on Android users' understanding of an app's data encryption and deletion practices. 
We examined whether Android users could locate this information on the label by asking if they thought they could request their data be deleted from Tumblr (the correct answer is ``No''). Then, we asked them whether knowing this information would impact their decision to download the app. Considering that the iOS privacy label does not provide information about an app's data encryption and deletion practices, we sought the opinion of iOS users on whether they would like such information to be included on the label. Additionally, we asked both Android and iOS users whether they would be less likely to download an app if they knew it did not encrypt their data or allow them to request data deletion from the app.


\paragraph{Final Impressions.} At the end of the interview, we instructed participants to stop screen sharing and asked them to reflect on their experience. We asked whether they would look at the privacy label when deciding to download an app in the future. Finally, we asked for participants' feedback on the privacy labels and provided them with an opportunity to add comments and ask any questions they might have.

\subsection{Data Analysis}
We automatically generated interview transcripts from our audio recordings using Otter.ai and Zoom and then manually reviewed and cleaned the transcripts. We inductively coded~\cite{cooper2012apa} our interview data. Two researchers read through the interview transcripts and collaboratively built a codebook. They independently coded one of the interviews and compared their codes. The entire research team examined the codebook together and discussed potential ambiguities in defined codes and made revisions accordingly. The two researchers who developed the codebook then each coded half of the remaining interviews and met regularly to discuss any issues identified by either researcher. They then reviewed the transcripts and decided on the appropriate coding together. The coders also met with the rest of the research team to discuss findings and update the codebook. 

\subsection{Limitations}
While we  used purposive sampling to ensure a diverse sample, our limited sample size of 24 participants is not representative of all iOS and Android phone users in the United States. Our sample skews more educated than the general U.S. population. Non-white racial and ethnic groups were underrepresented in our sample~\cite{US-demographics}. Recent work suggests that results from Prolific are reasonably representative of the U.S. population with regards to questions about privacy and security perceptions and experiences but not knowledge~\cite{tang2022replication}. Finally, we only recruited iOS and Android mobile phone users and did not explore other devices (e.g. tablets, laptops, smart watches) in Apple or Google’s ecosystem. User behaviors are likely to be different if they interact with the privacy label on a different device with a different interface layout. 

Due to the qualitative nature of our work, our coding is influenced by the attitude and interpretation of the researchers. Another group of researchers might identify different themes in our qualitative data. 

Even though we tried to mitigate bias by not mentioning anything related to privacy in our Prolific recruitment posts, it is still possible that participants learned the purpose of our study through the interview process and expressed more concern towards data practices than they otherwise would have.

Given the between-subjects design, any inconsistencies between the iOS label and Android label for our chosen apps might introduce a confounding effect. While Just Dance Now and Google Maps had consistent labels across both platforms, the Tumblr app had some notable inconsistencies. When we conducted our interviews, Tumblr did not report collecting Location in its Android label even though its iOS label reported Location as ``Data Used to Track You'' and ``Data Linked to You''; since then, Tumblr's Android label has been updated to include Location in the Data Collected section. Tumblr's Android label also indicated that ``no data was shared with third-party parties'' while its iOS label reported four data categories as ``Data Used to Track You.'' This apparent inconsistency might be due to differences in definitions (e.g., because this data use falls under one of the data sharing exclusions mentioned in Appendix~\ref{sec:sharing_exclusions}) or it might be an additional inconsistency between the two labels. 

\subsection{Ethical Considerations}
We took steps to ensure that this work followed ethical best practices. We recorded interviews with video off and avoided collecting identifiable information. We compensated all study participants for their time at a minimum rate of \$15 per hour. This study --- including interview scripts, recruitment strategies, and data practices --- was reviewed and approved in advance by the Carnegie Mellon University Institutional Review Board (IRB).

\section{results}
\label{sec:results}
In this section, we report the results from our user study. We first summarize participant demographics and present our findings related to overall user impressions and trust. We then examine the effect of each of the five key design differences outlined in Section~\ref{sec:background}, with our findings structured accordingly.


\subsection{Participants}
\label{subsec:participant_demographics}
Our non-representative U.S. sample includes participants with diverse backgrounds with respect to age, gender, ethnicity, and technology experience. 
Half of our 24 participants were Android users (denoted with A and a participant number), and half were iOS users (denoted with I and a participant number). 12 participants identified as male, 11 identified as female, and 1 identified as non-binary. 
The most common self-reported race or ethnicity was ``White or Caucasian'' with 13 participants selecting that option. 12 participants had at least a bachelor’s degree or equivalent, with five of those individuals reporting having obtained a master's degree. Two participants (both iOS users) had a degree in Computer Science or a related field. All participants downloaded at least one app in the past year.

At the beginning of our interview, we asked participants the sources they would consult to obtain information about the data collected by an app and its usage. None of the Android users mentioned privacy labels as a potential resource, while only two out of the 12 iOS users indicated they would refer to privacy labels in the Apple App Store. 

We present detailed information about  participant  demographics in \autoref{tab:demographics} in Appendix~\ref{sec:appendix_dem_survey}.

\subsection{Overall Impressions and Trust}

We observed similar overall impressions between Android and iOS labels. 
While many participants found the labels informative, some suggested that they were too vague and could be ``more detailed'' (I3). Others thought that ``they're probably being as, you know, transparent as they can'' (I5). About half the users in both groups described app privacy labels as helpful, but Participant A2 expressed a common sentiment:``It's helpful, but it needs a lot more help.'' Only one Android participant and five iOS participants recalled having previously seen a label. 11 Android participants and all 12 iOS participants claimed that they would look at app privacy labels again in the future.  

Impressions of the expanded labels were mixed. Some people found the additional details in the expanded labels informative. For example, Participant A3 said, ``it's a little bit more transparent with why they're collecting the data... I did appreciate what everything entailed...you could see reasons why, if you hit like the click down boxes and it provided more details as to why things like your approximate location and purchase history are being shared.''  Participant I2 noted, ``I like that it goes a little more in depth of what exactly they're tracking under each kind of bullet. So I kind of get more transparency of exactly what under those umbrellas are being tracked.''
Other participants wanted to see even more details about what types of information would be collected and how it would be used, and some noted being unfamiliar with some of the terminology used. Participant A11 described the expanded label as ``very vague of what to expect." Participant I5 explained, ``People are pretty skeptical when it comes to the honesty of companies. So I think [the expanded label is] a little more honest. But again, it's also a little concerning, because it's still pretty vague.'' 

Some participants had nuanced views about which data practices needed more or less detail.  For example, Participant I11 explained, ``Some of this information didn't really need to be expanded, like diagnostics, I can kind of, like, infer what they mean by that, or purchases, or maybe identifiers. But there were some other things that I appreciated maybe a little bit more [information about] like user content, I wouldn't have known what they meant at first.'' Multiple participants appreciated more information about iOS's ``User Content'' because they found the icon confusing. For example, participant I9 observed that the icon for that data category ``looks like the icon for like photos or whatever, so I wouldn't really want some random dance app having access to my photos.'' Other participants would have liked more detailed information about location. Participant A4 wondered, ``Like for location, does it go down to like at the level state level like do they know you're in California do they know you're in the county or do they know [or] guess what city you're in or even down to the precise coordinates?'' 



We did not observe any differences in beliefs about trustworthiness between Android and iOS labels. However, we observed that users' prior experience with an app and their understanding of its functionality can lead to confusion and concerns about the types of data collected. For example, participant I11, who used Google Maps once a week, questioned ``where money would come into play'' upon seeing that financial info was collected in Google Maps. Similarly, participant I10, who used Google Maps occasionally, expressed that she was surprised to see browsing history and financial info since she ``wouldn't assume that a map app would need that.'' When we asked participants who they thought created app privacy labels, most correctly identified that labels were created by app developers. However, eight of our 24 participants incorrectly believed that that the platform provider (Google or Apple) created the label. 

Many participants were skeptical about the accuracy of information reported in privacy labels, with 10 Android users and eight iOS users stating that they did not believe labels were completely accurate. Some participants mentioned concerns about developers fully disclosing their data practices. For example, participant A2 described themselves as 80\% confident about label accuracy ``because most companies are not forthright. Because there may be other sensitive information they collect but they just don't want to tell you.'' Indeed, researchers have found that app privacy labels are not always consistent with actual data practices~\cite{koch2022keeping,xiao2022lalaine,mozilla_2023}.

After reading the exclusions to definitions of data collection groups, some participants felt that the data presented came with caveats and loopholes that precluded accurate disclosure. For example, participant A1 mentioned that, ``they're playing a little loose with [the definition of data collection].'' 
%


\subsection{Location and Formatting of Links}

The location and formatting of links is different in the two labels. We observed that the link to the developer privacy policy at the top of the iOS label appeared to be more readily noticed than the links near the bottom of the Android labels.


We asked participants to visit the app store page for the Just Dance Now app and scroll to the label. We asked them to interact with the compact label and we observed whether they clicked on ``See Details'' to go to the expanded label, prompted only with the instructions that they should ``play and click around'' with the label. 10 out of 12 Android participants and seven out of 12 iOS participants navigated to the expanded label without being explicitly directed to click on the ``See Details'' link.

When we asked participants where they would go to search for more information if they were curious about what was present in the label or wanted to clear up any confusion, we received fairly different responses from participants on each platform.
Six iOS participants said they would look for information in the developer's privacy policy, while three said they would contact Apple and the rest said that they would search online using Google to find more information. In contrast, nine Android users said they would search online for more information and only one user said they would click on the developer's privacy policy link on the expanded label. None of the Android users mentioned referring to the ``Learn more'' link, which appears at the top of the expanded label and sometimes on the compact label if the app does not share data with third parties.
This difference may be due to the fact that the link to the developer's privacy policy appears prominently at the top of iOS labels (both compact and expanded) and appears only at the very bottom of Android expanded labels where users may be less likely to notice it.

\subsection{Data Collection Groups}

We asked all our participants to read the data collection group labels and explain their initial impression of their meaning. We defined a metric for correctness, incorrectness, and partial correctness for each term and evaluated participants' responses against the definitions provided by Google and Apple.

When we asked Android participants about the meaning of ``Data Shared with third parties,'' nine out of 12 provided a correct understanding of the definitions while three were only partially correct, focusing only on the sale of data rather than any type of sharing. Some participants also commented on the ambiguity of ``third parties,'' for example, participant A7 noted in their partially correct answer, ``I would wonder who are those third parties. Is this app getting paid to share them with somebody I don't know, that I probably don't want my data shared with. I have absolutely no idea.'' 

When we asked participants to explain the meaning of ``no data shared with third parties,'' eight provided correct definitions. For example, participant A8 said, ``I think they collect my data, but I don't think they're sharing it or selling it to third parties, for you know marketing purposes or whatnot.'' In addition, three participants provided partially correct definitions that focused on whether data was collected rather than shared. The final participant discussed data being used for advertising but did not offer a definition to answer our questions. 
After providing definitions, three participants said they were surprised that the Google Maps app did not share data and believed there would be sharing regardless of this statement in the compact label. 

When asked about ``Data Collected,'' eight out of 12 Android participants correctly defined ``Data Collected,'' with four offering definitions that were partially correct, misunderstanding where the data was stored (they said it was locally stored, but the Android definition of data collected focuses on data transferred off a user's device).

Consistent with previous work~\cite{zhang2022usable}, one of the most confusing terms for iOS users was ``Data Used to Track You,'' which only one participant was able to define correctly. I8 correctly defined tracking and stated: ``I think that the data is used to maybe recommend other apps and websites that I may like, and recommend and influence other purchases that I've made, I think, that is just being used to give me more recommendations and other things that I may be interested in.''
Participant I4 offered a partially correct response regarding tracking, stating that it is related to collecting information about their app activity and sharing the information with other companies (but without mentioning advertisers): ``I think that means like kind of keeping a log or tally of things that I do on the Internet, and then you know sharing that with other companies.'' Five users associated tracking with collecting usage data. For instance, participant I9 explained, ``I think they just want to know like screen clicks or how often the app is accessed.'' Three participants incorrectly thought tracking referred to location tracking across apps. 

When asked about ``data linked to you,'' eight iOS participants defined the term correctly, understanding that linking referred to connecting personal information with their identity. The remaining participants either did not understand linking or did not mention personal information or personal data being able to be linked. 
11 participants correctly defined ``data not linked to you.'' Participant I1, who misunderstood the term, stated that ``Data not linked to you, means that they're trying to tell me that they don't collect this information.'' 

Overall, participants had fairly accurate ideas about what was included in most of the data collection groups. However, we found significant confusion among iOS  users about ``Data used to track you,'' which some users confused with location tracking, web tracking, or collecting usage data.

\subsection{Completeness of Compact Label Groups and Category Lists}

The designers of both types of labels chose to leave out some information in the interest of keeping compact labels short. iOS labels omitted empty data collection groups while Android labels shortened long lists of data type categories. We found evidence that these design decisions impact what users take away from the compact labels.

We found many iOS users were unsure how to interpret the omission of a data collection group from a label. In order to assess their perception, we asked iOS participants whether they believed Google Maps was using their data to track them and if it collected data not linked to them. Even though the Google Maps iOS compact label did not include the ``Data Used to Track You'' and ``Data Not Linked to You'' boxes, 10 out of 12 users believed Google Maps was tracking them, while five believed it collected data not linked to them. For example, participant I8 said she thought Google Maps collected data not linked to her because the label didn't have a section definitively stating that no such data were collected; she said she would like to see ``a clear, bold statement that no other data is being collected... just have a section that specifically [says] that nothing else is going to be collected.''

Several Android participants asked about the elision of data type categories in the compact label---for example, participant A7 asked, ``What are the other nine things they're collecting?''---and some criticized it. For example, when asked what changes they'd like to make to the label, participant A8 said, ``I'd like to know the nine other data types that they're collecting.'' 
When we asked participants about what data types they cared about most in the compact labels, seven Android users mentioned financial information, three mentioned location, and three mentioned personal information. Due to the way the Android label abbreviates the list of data types in the compact label, these three data types are the ones that most commonly appear without clicking through to the full label, so these were likely the most salient data types for our participants at the time we asked this question. 

\subsection{Expanded label structure}

The iOS expanded label focuses on data collection purposes (with data type categories listed under each purpose), while the Android expanded label focuses on data type categories (with purposes listed under each data type). We found that both designs are confusing to users, albeit in different ways.

To explore iOS users' understanding of the data collection purpose groupings in the expanded label, we instructed participants to look at the three data uses in the Just Dance Now expanded label under ``Data Linked To You''---``Analytics,'' ``Product Personalization,'' and ``App Functionality''---and asked them to explain why there were three groups shown. Half of the iOS participants showed confusion towards the groupings and the other half were able to correctly identify the groups and why they were structured in the way they were. 
When iOS participants viewed the Google Maps label, we asked them about any differences between this label and the previous label (Just Dance Now) that jumped out. Four participants noticed and commented on data being used for third-party advertisements while two participants commented on data being used for analytics. None of the other participants appeared to notice the differences in how data was used between the two apps. 

Android labels hid their purposes behind an accordion interface that required participants to click on the chevron symbol next to each data category to find out how that type data was used. 
Seven Android participants clicked to reveal at least one data category unprompted. 
Of these, two participants revealed one category, four participants revealed between four and seven categories, and one participant revealed all 11 categories. This suggests that most Android users who view the expanded label are likely to get an overview of the types of data collected but only a glimpse of some of the purposes for which collected data is used. 
The lack of visibility into data use purposes became apparent when we asked Android users to view the expanded Tumblr label and tell us whether any data types were used for advertising or marketing purposes. The expected response was ``Device and other IDs,'' which Android participants could find by navigating through individual data type categories and reviewing the corresponding collection purposes in the accordion interface. Only two of 12 participants successfully located ``Device and other IDs,'' while the remaining participants did not know where to find the data type and provided incorrect answers.

Our observations of both Android and iOS users' failures to notice information about data use purposes suggest that neither approach to structuring expanded labels is all that successful. While the iOS approach is purpose-centric, many users did not seem to understand that they were being shown purpose-related data groups. On the other hand, Android users seemed not to recognize that they needed to expand the accordion in order to reveal purpose information.

\subsection{Representation of Data Security Practices}

Android Data Safety Labels include information about data security practices in addition to information about data collection and use. All Android labels show whether users can request their data be deleted and whether the app encrypts data in transit. The label also displays if the app has undergone an external security review. Android participants were generally able to offer reasonable explanations of these terms and found this information useful; iOS participants indicated that they would find such information valuable. 

\paragraph{Data Deletion} Ten participants correctly defined ``Data deleted,'' with the remaining two participants offering partially correct definitions that suggested that third parties could still access deleted data. For example, participant A10 explained, ``Yeah I think that any third parties that's not part of the company that I requested it be deleted from would still have access to it.'' 
When discussing data deletion, some participants noted that they interpreted this as a ``soft'' deletion in which the company will still have access to the data in their records and can still use it as needed. For example, Participant A8 observed, ``I'm imagining that it just deletes my account on my account information, but I don't know if the company still has the information saved somewhere else, like on another server or database or whatnot.'' A4 mentioned that deletion behavior might depend on local regulations. He  explained, ``It just depends on where you're at before I guess. For me, since I'm in California, I would say yes, I do think it's deleted. But if I was like in a neighboring state, I don't think it would.''

Several participants found the information on data deletion particularly salient: Participant A4's first impression of the Tumblr app was, ``It shows data can't be deleted,'' and when asked for their overall impression of Tumblr, Participant A3 said ``I definitely don't like the last point with data can't be deleted.'' Nine Android participants also said they would be less likely to download Tumblr because it did not allow users to request that data be deleted. Seven out of 12 iOS participants  also said that they would be less likely to download an app if it did not allow their data to be deleted. Some users who said they would still download Tumblr commented that it might affect other behavior. For example, I7 commented, ``I may still download it, but it definitely affects how much I used it probably. I would probably still use it to view other's posts but I probably wouldn't post anything myself.'' 

When we asked iOS users whether they would like to be informed about an app's data deletion practices on the privacy label, 11 out of 12 users said that it would be helpful, with Participant I11 stating, ``Of course, that's always something that I think people like me could benefit from.'' The iOS participant who said ``no'' thought that the inclusion of an app's data deletion practices might not be necessary on the privacy label but they would like to see ``a pop-up as a warning when [the company] does delete their data.'' iOS users expressed similar concerns to those of Android users with regards to the commitment to data deletion requests, as I6 commented, ``It would be nice if I could press a button and just delete everything I ever put on Facebook, but I don't think that's possible and I don't think they would do it.''

\paragraph{Data Encryption} Encryption was less widely understood than other terms, with only four participants able to provide fully-accurate definitions of ``Data Encrypted,'' (e.g., A8 said, ``it's like garbled and unable to read''). Participant A5 provided a partially correct answer, mentioning HTTPS, a protocol used for encrypted data transfer: ``Now I believe data encrypted means that it’s not set out without https, so it should be encrypted where it can’t be broken into in transit.... 
Between points so that it can't easily be stolen by a hacker or someone else.'' Other participants confused encryption with compression (A4) or anonymization (A10), stated that they did not know, or offered vague answers. However, most participants understood that encrypting data in transit was a desirable feature for apps, and it was generally expected. For example, A4 commented, ``The data is encrypted in transit: that's probably pretty standard now.''

Android users indicated that whether apps used encryption was important to them. 
Ten out of 12 Android participants said they would be less likely to download an app if it doesn't encrypt their data in transit. Participant A4 commented, ``If they don't take something as like a simple as encrypting my data in transit, I wouldn't be able to trust them with something much more major like holding my my credit card information or my messages.'' 

iOS users also indicated that information about encryption would be a valuable addition to iOS App Privacy Labels. Seven out of 12 iOS participants reported that they too would be less likely to download an app if it did not encrypt data in transit. Reasons provided included that they wanted to prevent data loss due to a data breach or have their data be easily accessible by malicious third parties or companies. 
Participant I5 said if data was not encrypted, ''it just gives [hackers] an easier time getting your information....'' 

\paragraph{Independent Security Review} We asked participants what they thought was meant by ``independent security review'' on the Google Maps label.
Most participants seemed to understand generally what this term referred to. 
Nonetheless, some participants were skeptical about the idea of an independent third party performing the review and wondered whether they would 
be neutral in their assessment. However, a majority of  participants found this information to be useful and said it increased their trust in the app. 
Participant A9 pointed out this raised their trust and confidence in the app since ``[the independent security review] shows that Google is confident and lets third parties audit their product.'' 

Information about a security review was highly salient; when participants viewed the compact label for Google Maps and were asked to compare it with the previous label, they all noticed that, unlike Just Dance Now, Google Maps had an independent security review and did not share data with third parties.
Despite some uncertainty about exactly what an independent security review is and who conducts it, 11 Android users reported it to be helpful. For example, Participant A8 said, ``It makes me feel like it's safer, whether or not that's true is another story.'' Eight Android users thought that there might be risks to downloading an app that had not been independently reviewed, with A8 asking, ``If a company doesn't have it now that I know that it exists, I am like why don't they have that?''

We also asked iOS users whether having an independent security review of Google Maps would make the app more secure and boost their confidence in its use. Nine out of 12 of them responded positively. However, a majority of them also expressed concerns about the reviewing entity's affiliations with the company that owns the app. For example, participant I5 mentioned that their confidence in Google Maps' security would be higher if the term ``independent'' truly implied a complete third party with no affiliations.


\section{Discussion}
\label{sec:discussion}

Our results suggest that while privacy labels offer an opportunity to enhance transparency about mobile apps' data practices, the design of both iOS App Privacy Labels and Android Data Safety Labels could be improved to further enhance privacy. 

\begin{rec} Privacy labels should use clear, comprehensible terms. \end{rec}

Like prior work~\cite{zhang2022usable,tang2021defining}, we found that many people were unable to correctly identify the term ``track,'' with many iOS users associating the term with location tracking. Other users primarily understood tracking as data collection, leading to a lack of clarity about the differences between the three data collection groups in the iOS label and consequent confusion when trying to comprehend these groups and their implications. The term ``Data Linked to You'' was also poorly understood. Although Android's distinction between ``Data Shared'' and ``Data Collected'' seemed more intuitive to people, most user definitions failed to account for exclusions defined by the Google Play Store (Appendix~\ref{sec:appendix_google_exclusions}); these exceptions might lead to misinterpretations of the disclosures reported in an Android Data Safety Label. 

Some users also mentioned that they weren't familiar with the terms used to describe data type categories and data types. For example, after looking at an expanded Android label, participant A10 commented, ``It's more information I don't really know a lot about it. I think it would depend heavily on the users like experience and knowledge about what these words mean, whether or not it meant anything to them.'' Use of unfamiliar or misunderstood terms necessarily undermines labels' ability to inform users about data practices.

App stores and usable privacy researchers should conduct user studies and usability testing to form an empirical basis for selecting terminology to include in future privacy labels. Further work will be required to identify terms that clearly capture the described uses and data types and to explore how technical terms might be augmented with explanatory language. This work should also ensure that users can clearly distinguish different groups and categories. Privacy label designers should carefully consider the placement of links that offer extended privacy information and ensure that the positioning encourages user engagement.

\begin{rec}Groups and categories should not conflate elements with different privacy implications.
\end{rec}

In several cases, participants exhibited confusion about data type categories that suggested these categories were too broad. For example, ``User Content''---a term used in iOS labels---appeared to participant I9 to include user photos because it was accompanied by a photo icon. I9 commented that she didn't think a dance app should have access to her photos. 
However, in that label, the ``User Content'' listed under ``Data Used to Track You'' referred exclusively to customer support data, whereas the ``User Content'' listed under ``Data Linked to You'' additionally included photos or videos. The participant felt differently about photo usage compared to customer support data, but both types were conflated under the same data category. Similarly, Android users thought the term ``Financial Info'' was too vague and they didn't know what it contained; in the expanded label of Just Dance Now, it only included purchase history, but there were concerns that it could include credit card information. Participants felt differently about sharing purchase history compared to sharing credit card information. 

To enhance transparency, labels should avoid using ambiguous category terms---especially those that include (or appear to include) individual types with significantly different privacy or security implications. Iterative user studies should be conducted to identify groups and categories that avoid conflating individual elements with different privacy implications. 

\begin{rec} Privacy labels should use clear, concise design elements to convey critical information about data practices.     
\end{rec}

Two pieces of information that users  consistently identify as critical to their privacy assessment include (1) what information is collected and (2) the purpose that information is used for. In both cases, iOS and Android labels present that information differently, and in both cases, existing designs leave room for improvement. 

In most cases, the compact Android label lists only a subset of the data type categories collected. For example, the label for Just Dance Now states that the app may collect ``Location, Personal Info and 4 others'' whereas the Google Maps label states that it may collect ``Location, Personal Info and 9 others.'' Some Android users expressed confusion and suspicion about the information elided under ``and $n$ others.'' In contrast, iOS labels include a full list of data type categories under each of their data collection groups, more clearly conveying information about data collected while retaining a relatively concise design.

Despite the availability of data collection purposes within both Android and iOS labels, users faced challenges in understanding the information. Only two out of 12 Android users answered correctly when asked about the specific data types collected for a given purpose. Similarly, among iOS users, only half could provide an explanation for the presence of three data use groups. These findings indicate that users encountered difficulties in locating information about the purpose of data collection, and users who did locate it struggled to comprehend the provided information. These findings highlight the need for a clearer, concise representation of data collection purposes in privacy labels. A tabular or matrix format has been suggested as a compact and intuitive solution in previous research~\cite{kelley2009nutrition,kelley2010standardizing, reinhardt2021visual}. In the case of mobile app labels a matrix may be especially well-suited for representing the interplay between data types and purposes. While adapting this format for mobile devices may pose design challenges due to limited screen size, it seems worth attempting to design a matrix format that will work on a small screen.


\begin{rec} Privacy labels should explicitly include absent data collection categories.     
\end{rec}


There exists a notable disparity between the Android and iOS privacy labels in how they represent the absence of collected or shared data: Android Data Safety labels explicitly state when no data types are collected, whereas iOS App Privacy Labels simply omit the box for the relevant data collection group. For example, the Android privacy label for Google Maps states ``No data shared with third parties'' whereas the iOS label simply omits the ``Data Used to Track You'' and ``Data Not Linked to You'' boxes. We found that iOS users misinterpreted the omitted information: 10 out of 12 believed that Google Maps tracked them and five out of 12 believed it collected data not linked to them. By contrast, there was no confusion about interpreting the statement ``No data shared with third parties.'' Similarly, Android users noticed that the Tumblr app did not include an option to request to delete data; we suspect that simply omitting that information would not have emphasized that information to users in the same way. 

To address user confusion and misinterpretation, privacy labels should include explicit text indicating when no data are collected. Further research should explore whether there are other omissions that should be explicitly included, for example, users might appreciate a label that emphasizes when no location data is collected or when no security review has been conducted. A matrix representation may be able to aid in showing both the presence and absence of particular data practices.


\begin{rec} Privacy Labels should include information about security practices. 
\end{rec}

Unlike iOS labels, Android labels include information about two security practices: whether data are encrypted in transit and whether a user can request to delete their data. We found that this information---especially when security practices were not adopted---impacted user trust and download intentions. Additionally, if an app has undergone an independent security review, the Android label stated so. Nearly all (11 out of 12) Android users found this information helpful, despite some uncertainty about the reviewer's identity. One person commented, ``It's good to know that their security is reviewed...you can have a little more faith [in the app] if you know that they're being reviewed by someone who's not part of their business.''

Based on our results, we recommend that privacy labels include information about security practices in addition to information about data collection and use. For apps that have undergone an independent review, the label should clearly identify the reviewing entity and clarify any connections they might have with the app's company. However, we note that the information included in current Android labels does not necessarily capture all relevant security practices. Future work should explore whether there are other security practices that should be included in privacy labels and how to best express security practices in a way that users can understand.

\begin{rec}App stores should automatically verify privacy labels.     
\end{rec}

Our results highlight that while many Android and iOS users were uncertain about the entities responsible for creating app privacy labels, the majority (16 out of 24) could correctly identify app developers as the label creators. These users expressed skepticism regarding the accuracy of the privacy labels. This skepticism is validated by prior work, which has found inconsistencies between privacy labels and app data practices~\cite{koch2022keeping,li2022understandingB,xiao2022lalaine}. 

To enhance users' trust in the accuracy of privacy labels, we recommend that app stores introduce tools to help developers create accurate labels~\cite{li2021honeysuckle,gardner2022helping} and require automated privacy label verification as part of the app update process. Prior work has shown that statistical analysis can be used to identify discrepancies between reported data use practices and actual practices~\cite{koch2022keeping}. Further work will be required to reliably and accurately verify the full contents of app privacy labels. 

We also recommend that privacy labels clearly identify the principal responsible for creating the label along with information about how the label was (or was not) verified. 

\section{Conclusion}
\label{sec:conclusion}
Although privacy labels increase transparency of mobile app data practices, our usability study sheds light on the difficulties faced by Android and iOS users in understanding the information presented in privacy labels. Findings from our research provide valuable insights into the impact of Android and iOS privacy labels on user comprehension, as well as their influence on the decision-making process when downloading an app. Additionally, we identify design issues associated with these labels, forming the basis for concrete recommendations to refine and improve their effectiveness.

\begin{acks}
This research was supported in part by the Innovators Network Foundation, Google, and the National Science Foundation under grant CNS-2150217. 
\end{acks}

\bibliographystyle{ACM-Reference-Format}

\appendix
\label{sec:appendix}

\definecolor{Gray}{gray}{0.9}
\begin{table*}
    \begin{threeparttable}
    \caption{Android \& iOS Participant Demographics}
    \label{tab:demographics}d
    \centering
    \begin{tabular}{ c|c|c|M{50mm}|M{50mm}|c } 
     \hline
     PID & Age Group & Gender & Ethnicity & Education & Tech Exp\\
     \hline 
     \rowcolor{Gray}
     A1 & 42 - 49 & Male & White or Caucasian & Some college but no degree & No\\ 
     A2 & 50 - 57 & Female & Black or African American & Some college but no degree &  No\\ 
     \rowcolor{Gray}
     A3 & 26 - 33 & Male & Asian & Master's Degree & No\\ 
     A4 & 26 - 33 & Male & Latinx or Hispanic & Master's Degree & No\\ 
     \rowcolor{Gray}
     A5 & 34 - 41 & Male & White or Caucasian & Master's Degree & No\\ 
     A6 & 26 - 33 & Male & White or Caucasian & Some college but no degree & No\\ 
     \rowcolor{Gray}
     A7 & >57 & Female & White or Caucasian & Bachelor's degree in college (4-year) & No\\ 
     A8 & 34 - 41 & Female & White or Caucasian & Associate degree in college (2-year) & No\\ 
     \rowcolor{Gray}
     A9 & 18 - 25 & Male & Black or African American & High school graduate (or equivalent) & No\\ 
     A10 & 34 - 41 & Female & Native American  \& Latinx or Hispanic & Some college but no degree & No\\ 
     \rowcolor{Gray}
     A11 & 18 - 25 & Female & Latinx or Hispanic & Some college but no degree & No\\ 
     A12 & 42 - 49 & Non-Binary & White or Caucasian & Bachelor's degree in college (4-year) & No\\ \rowcolor{Gray}
     I1 & 50 - 57 & Female & White or Caucasian & Some college but no degree & No\\ 
     I2 & 26 - 33 & Male & White or Caucasian & Bachelor's degree in college (4-year) & No\\ 
     \rowcolor{Gray}
     I3 & 26 - 33 & Female & Asian & Bachelor's degree in college (4-year) & No\\ 
     I4 & 34 - 41 & Male & Prefer not to respond & Bachelor's degree in college (4-year) & No\\ 
     \rowcolor{Gray}
     I5 & 18 - 25 & Male & White or Caucasian & Bachelor's degree in college (4-year) & No\\ 
     I6 & 34 - 41 & Female & White or Caucasian & Master's Degree & No\\ 
     \rowcolor{Gray}
     I7 & 42 - 49 & Male & White or Caucasian & Associate degree in college (2-year) & Yes\\ 
     I8 & 18 - 25 & Female & White and Black & Bachelor's degree in college (4-year) & No\\ 
     \rowcolor{Gray}
     I9 & 42 - 49 & Female & White or Caucasian & Master's Degree & No\\ 
     I10 & 26 - 33 & Female & Mixed black and white & Some college but no degree & No\\ 
     \rowcolor{Gray}
     I11 & 18 - 25 & Male & Asian & Associate degree in college (2-year) & No\\ 
     I12 & 26 - 33 & Male & White or Caucasian & Associate degree in college (2-year) & Yes\\ 
     \hline
    \end{tabular}
         \begin{tablenotes}
          \small
          \item ``PID'' refers to participants' identifiers, the PID of an Android user starts with ``A'' and 
          the PID of an iOS user starts with an ``I.'' ``Tech Exp'' refers to whether the participant has a degree in Computer Science or a related field.
        \end{tablenotes}
    \end{threeparttable}
\end{table*}

\section{Demographic Survey}
\label{sec:appendix_dem_survey}
\begin{flushleft}
1. What is your Prolific ID? This will help us pay you through Prolific. [free-text]\\
2. What type of device do you currently own?\\
\-\hspace{0.2cm} a. Android\\
\-\hspace{0.2cm} b. iOS\\
3. What is the brand and model of your phone? (For example, ``Samsung Galaxy S22'') [free-text]\\
4. Please select the age group you are in.\\
\-\hspace{0.2cm} a. 18 - 25\\
\-\hspace{0.2cm} b. 26 - 33\\
\-\hspace{0.2cm} c. 34 - 41\\
\-\hspace{0.2cm} d. 42 - 49\\
\-\hspace{0.2cm} e. 50 - 57\\
\-\hspace{0.2cm} f. Above 57\\
\-\hspace{0.2cm} g. Prefer not to respond\\
5. What gender do you identify with?\\
\-\hspace{0.2cm} a. Male\\
\-\hspace{0.2cm} b. Female\\
\-\hspace{0.2cm} c. Prefer to self-describe: \textunderscore\\
\-\hspace{0.2cm} d. Prefer not to respond\\
6. How do you describe your race or ethnic identity? (You may select more than one option.)\\
\-\hspace{0.2cm} a. American Indian or Alaska Native\\
\-\hspace{0.2cm} b. Latinx or Hispanic\\
\-\hspace{0.2cm} c. Asian\\
\-\hspace{0.2cm} d. Native Hawaiian or Pacific Islander\\
\-\hspace{0.2cm} e. White or Caucasian\\
\-\hspace{0.2cm} f. Prefer to self-describe: \textunderscore\\
\-\hspace{0.2cm} g. Prefer not to respond \\
7. What is the highest level of education you have received? \\
\-\hspace{0.2cm} a. Less than high school degree\\
\-\hspace{0.2cm} b. High school graduate (high school diploma or equivalent including GED)\\
\-\hspace{0.2cm} c. Some college but no degree\\
\-\hspace{0.2cm} d. Associate degree in college (2-year)\\
\-\hspace{0.2cm} e. Bachelor's degree in college (4-year)\\
\-\hspace{0.2cm} f. Master’s Degree\\
\-\hspace{0.2cm} g. Doctoral Degree\\
\-\hspace{0.2cm} h. Professional degree (JD, MD)\\
\-\hspace{0.2cm} i. Prefer not to respond\\
8. What is your annual household income?\\
\-\hspace{0.2cm} a. Less than \$25,000\\
\-\hspace{0.2cm} b. \$25,000 - \$50,000\\
\-\hspace{0.2cm} c. \$50,000 - \$100,000\\
\-\hspace{0.2cm} d. \$100,000 - \$200,000\\
\-\hspace{0.2cm} e. More than \$200,000\\
\-\hspace{0.2cm} f. Prefer not to respond\\
9. Do you hold a job or degree in computer science or a related field?\\
\-\hspace{0.2cm} a. Yes\\
\-\hspace{0.2cm} b. No\\

\end{flushleft}

\section{Android Interview Scripts}
\label{sec:appendix_android_scripts}
\subsection{Introduction}
\begin{flushleft}
– Thank you for meeting with me today. This interview is being conducted for research at [ANONYMOUS INSTITUTION] to better understand how people interact with mobile apps in the Google Play Store. I will ask you to answer some questions and view some information in the Play Store. This session should take around an hour to complete, and will be recorded via Zoom. You have previously received 40¢ for scheduling the interview, and upon completion of the study, you will be paid \$14.60 through Prolific. You will be asked to share your phone screen via Zoom at some point during the session to enable us to follow what you are doing on your phone as you visit the Play store.

– Please answer our questions truthfully and as thoroughly as possible. If in doubt, feel free to ask me for clarification at any point during the interview. I want to emphasize that there are no right or wrong answers. Our goal is simply to understand your opinions and thought processes. You may stop the interview at any point, or choose to not answer a question, or take a break if you wish. Please do not reveal any private or personally-identifiable information about yourself or others during the interview. If you accidentally reveal any personal information, please let me know so that I can remove it from the recording. You have already received the consent form and agreed to it. Do you have any questions at this time? \\

– Please turn off your camera so that your face is not recorded for the interview.\\
\end{flushleft}
\subsection{General Questions about App Usage}
\-\hspace{0.2cm}- When was the last time that you downloaded a new app on your phone? [Follow-up] Have you downloaded an app in the past year or so?\\
\-\hspace{0.2cm}- What are some of the typical factors that influence which apps you download on your phone? (Prompt: app reviews, brand, ratings, security, ranking, data privacy, your friends) \\
\-\hspace{0.2cm}- Have you ever compared different apps before deciding which one to download (Prompt: what types of things have you compared?)? \\
\-\hspace{0.2cm}- Have you ever considered data privacy in the decisions you make? Why/Why not?\\
\-\hspace{0.2cm}- Have you ever wondered what information/data apps collect about you?\\
\-\hspace{0.4cm}- [If they say they have] How would you go about finding out what information an app collects about you and what the app does with the information?\\
\-\hspace{0.4cm}- [Follow-up] Have you ever actually done that?\\
\-\hspace{0.4cm}- [If they say they have not] If you were to look for this, how could you possibly find out what information an app collects about you and what the app does with that information? Prompt: media? friends/family? experts? privacy policies/EULA? permission settings? Other? \\
\-\hspace{0.4cm}- Do you think those sources are reliable? Google Play store? Have you ever looked?\\
\-\hspace{0.2cm}- Is there any information in the Google Play Store you could turn to if you wanted to consider privacy in your app download decisions? [If they say yes, ask them what sort of information it is and where they would find it.]\\
\-\hspace{0.4cm}- [If they mention the privacy/data safety label] Do you know what kind of information you would find if you looked at one of these labels?\\
\-\hspace{0.4cm}- [If they don’t mention the privacy/data safety label] Google recently started including data safety labels for apps in the play store that provide a summary of privacy information. What sort of information would you expect this summary might contain?\\
- [Connect phone to Zoom] If you do not have Zoom on your phone already, please head to the Play Store and search for the Zoom app. Download the app on your phone and try to connect to the meeting by clicking on the Zoom link we sent to you earlier. There is no need to join the audio from your phone. There should be a green rectangle on the screen which will allow you to share your phone screen with us. We will be recording what is on your phone screen and thus recommend closing or hiding any apps that you would not like us to accidentally record. You could do so by turning on the ``Do Not Disturb'' feature on your phone. (A circle with a minus sign in the center for Samsung phones) \\
At this moment I would like to ask you to turn on ``Do Not Disturb'' from your device. You can do this by swiping down from the top and clicking the circle with a minus sign.\\
\subsection{Just Dance Now}
- [App Familiarity] Before we look at any labels, are you familiar with the Just Dance Now app?\\
\-\hspace{0.2cm}- Could you describe what this app does?\\
\-\hspace{0.2cm}- Have you used Just Dance Now before? \\
\-\hspace{0.4cm}- [If yes] how often do you use it? \\
\-\hspace{0.4cm}- [If no] It's an app for dancing to music with steps provided by the app on your phone. \\
- We will be taking a look at the privacy label of Just Dance Now. Could you navigate to the Play Store and search for Just Dance Now?\\
- Please scroll down to the ``Data Safety'' section, and now I would like to give you a moment to take a look at the info. \\
\-\hspace{0.2cm}- [Overall impression] What is your overall impression of this label? \\
\-\hspace{0.2cm}- Have you seen a label like this before?\\
\-\hspace{0.4cm}- [If yes] Where have you seen it? \\ 
\-\hspace{0.4cm}- [If yes] What did you do when you saw it?\\
\-\hspace{0.2cm}- [expectations] Is the privacy information you see here in line with what you would expect for an app like this?\\
\-\hspace{0.2cm}- Did you find anything that was surprising/unexpected?\\
\-\hspace{0.4cm}- [If they did] What surprised you about this?\\
\-\hspace{0.4cm}- [If they didn't] Why did you expect to see this information?\\
\-\hspace{0.2cm}- Is there other information about how this app uses data that you would like to know?\\
- [Label Comprehension]\\
\-\hspace{0.2cm}- [Data Sharing] What do you think the phrase ``This app may share these data types with third parties'' means?\\
\-\hspace{0.4cm}- [Clarification] What do you think an app does with this data?\\
\-\hspace{0.4cm}- [If they say the app will share your data] what does sharing mean in this context? \\
\-\hspace{0.4cm}- [Follow up if needed with:] How do you think sharing works?\\
\-\hspace{0.4cm}- Looking at the list of things under ``This app may share these data types with third parties'' does any of this concern you? If so, what? \\
\-\hspace{0.6cm}- Why does this concern you? \\
\-\hspace{0.6cm}- Why does that not concern you?\\
\-\hspace{0.2cm}- [Data Collection] What do you think the phrase ``This app may collect these data types'' means?/What does data collection mean?\\
\-\hspace{0.4cm}- [Clarification] What do you think an app does with this data?
[If they say the app will collect data on you] what does collecting mean in this context?\\
\-\hspace{0.4cm}- How do you think the app does that?\\
\-\hspace{0.4cm}- Looking at the list of things under ``the app may collect these data types'', does any of this concern you? If so, what? \\
\-\hspace{0.6cm}- Why does this concern you? \\
\-\hspace{0.6cm}- Why does this not concern you?\\
- [Data Encryption] What do you think the phrase ``Data is encrypted in transit'' means?\\
\-\hspace{0.2cm}- [Clarification] What do you think an app does with this data?\\
\-\hspace{0.4cm}- [If they say the app will encrypt data] what does encryption mean in this context?\\
\-\hspace{0.4cm}- [clarify. If the interviewee has a wrong idea of encryption] data encryption means that information is converted into secret code that hides the information and helps make sure that nobody can eavesdrop and see the information. Some apps use data encryption whenever they send information to the server so that eavesdroppers can’t read your personal information. \\
- [Data Deletion] What do you think the phrase ``You can request that data be deleted'' means?\\
\-\hspace{0.2cm}- [Clarification] What do you think an app does with your data when you make this request?\\
\-\hspace{0.4cm}- [If they say the app will delete your data] what does deletion mean in this context? [Permanent delete vs soft delete?]\\
\-\hspace{0.2cm}- Are apps required to delete your data if you request that data be deleted? \\
\-\hspace{0.2cm}- What do you think happens to the data after you request for it to be deleted?\\
{[Record whether the user has clicked on \textbf{``See Details''} by this point]}\\
\-\hspace{0.2cm}- [if yes] start asking specific questions below\\
\-\hspace{0.2cm}- [If not] ask ``do you see anywhere you can click to get more information about this app’s collection or use of data?'' \\
\-\hspace{0.4cm}- [Note if the user suggests other things they could click on] \\
\-\hspace{0.4cm}- [If they say ``See Details''] let them go ahead and click that\\
\-\hspace{0.4cm}- [if they don’t] suggest ``See Details'' and ask if they can find the ``See Details'' link\\

\subsection{Google Maps}

\-\hspace{0.2cm}- Now we are going to read the privacy label of a different app. \\
Are you familiar with Google Maps? 
Could you describe what this app does? Google Maps is a web-based service that provides detailed information about geographical regions and sites around the world.\\
\-\hspace{0.2cm}Have you used Google Maps before? \\
\-\hspace{0.4cm}- [If yes] how often do you use it? \\
\-\hspace{0.4cm}- [if not, explain:] \\

Ok, let’s now take a look at Google Maps's privacy label. Could you search for Google Maps in the Google Play Store? Please click on ``Data Safety'' to expand it. \\
\-\hspace{0.2cm}- [overall impression] What is your overall impression of this label?\\
\-\hspace{0.4cm}- Do any differences between this label and the previous label jump out?\\
\-\hspace{0.2cm}- [expectations] Is the privacy information you see here in line with what you would expect for an app like this?\\
\-\hspace{0.2cm}- Do you see any information that is surprising or unexpected?\\
\-\hspace{0.4cm}- [If they did] What surprised you?\\
\-\hspace{0.4cm}- [If they didn't] Why did you expect to see this information?\\
\-\hspace{0.2cm}- Is there other information about how this app uses data that you would like to know?\\

- [Label Comprehension]

\-\hspace{0.2cm}- What do you think ``No data shared with third parties'' means?\\
\-\hspace{0.2cm}- Do you find this surprising? \\
\-\hspace{0.4cm}- [if so] Why is this surprising? \\
\-\hspace{0.4cm}- [if not] Why is this not surprising?\\
\-\hspace{0.2cm}- What do you think ``independent security review'' means?
[clarification if not mentioned] Who do you think did the independent security review?\\
\-\hspace{0.2cm}- Do you find this information helpful?\\
\-\hspace{0.4cm}- [if yes] Why is this helpful?\\
\-\hspace{0.4cm}- [if no] Why is this not helpful?\\
\-\hspace{0.2cm}- Does this information make you feel more confident about the security of Google Maps?\\
\-\hspace{0.2cm}- Do you think there are any risks to downloading an app that hasn't had an independent security review? \\
\-\hspace{0.4cm}- If so, what do you think they might be?\\
Please click on \textbf{``See Details''} to expand the privacy label.\\
\-\hspace{0.2cm}- Are you surprised about anything you see in the expanded label? \\
\-\hspace{0.4cm}- [If so] Which information do you find surprising to see here?\\

\subsection{Tumblr - Fandom, Art, Chaos}
Now we are going to read the privacy label of a different app.\\ 
\-\hspace{0.2cm}- Are you familiar with Tumblr?\\
\-\hspace{0.2cm}- Could you describe what this app does? [Tumblr is a microblogging and social networking platform where users could post multimedia and other content to a short-form blog]\\
\-\hspace{0.2cm}- Have you used Tumblr before? \\
\-\hspace{0.4cm}- [If yes] how often do you use it?\\
\-\hspace{0.4cm}- [if not, explain:]\\

Could you please exit the current label and find ``Tumblr – Fandom, Art, Chaos'' in the Google Play Store? Please scroll down to go to the ``Data Safety'' section as usual or click on this link we sent you 
\-\hspace{0.2cm}- [overall impression] What is your overall impression of this label? \\
\-\hspace{0.4cm}- Do any differences between this label and the previous labels jump out?\\
\-\hspace{0.2cm}- [expectations] Is the privacy information you see here in line with what you would expect for an app like this?\\
\-\hspace{0.2cm}- Do you see any information that is surprising or unexpected?\\
\-\hspace{0.4cm}- [If they did] What surprised you?\\
\-\hspace{0.4cm}- [If they didn't] Why did you expect to see this information?\\
\-\hspace{0.2cm}- Is there other information about how this app uses data that you would like to know?\\
\textbf{Note whether the user has opened ``See Detail'' by now, if not, ask them where they could find more privacy information, ask them to click on ``See Details'' if they could not find it}\\
\-\hspace{0.2cm}- How securely do you think Tumblr handles your data?\\
\-\hspace{0.2cm}- [data encryption] Do you think Tumblr encrypts data in transit?\\
\-\hspace{0.2cm}- What do you think will happen if your data isn't encrypted in transit?\\
\-\hspace{0.2cm}- Who do you think could get access to the unencrypted data?\\ 
\-\hspace{0.2cm}- Would you be less likely to download Tumblr if it doesn’t encrypt data in transit?\\
\-\hspace{0.4cm}- Why / Why Not?\\
\-\hspace{0.2cm}- [data deletion] Do you think that you could request your data be deleted from Tumblr?\\
\-\hspace{0.4cm}- [if yes] Why do you think so?\\
\-\hspace{0.4cm}- [if not] What does ``data can’t be deleted'' mean to you?
Would you be less likely to download Tumblr since it doesn't allow you to request your data be deleted?\\
Why / Why Not?\\

\-\hspace{0.2cm}- [Advertising/Marketing] Is there any information collected here used for advertising or marketing purposes? \\
\-\hspace{0.4cm}- Record whether the user could find it under ``Device or Other IDs,'' if they could not find it themselves, direct them to ``Device or Other IDs''\\
\-\hspace{0.2cm}- [Missing data type – location] Do you think that Tumblr uses your location to personalize content/ads?\\
\subsection{Reading Definitions}

- If you want to learn more about what ``Data collected'' and ``Data shared'' mean, where would you go to find this information?\\
\-\hspace{0.2cm}- [If they missed it] Tell them that if they go to the expanded label, there is a ``Learn More'' link below ``Data Safety,'' go ahead and click on that. \\
\-\hspace{0.2cm}- [If they still couldn't find it] send a link to the label in the Zoom chat\\
\-\hspace{0.2cm}- Scroll down a little bit and take a moment and go over the definitions of ``Data collection'' and ``Data sharing'' under ``Understand \& review app data safety practices.'' Please let me know when you have finished reading them. \\
\-\hspace{0.2cm}- Is the definition of ``data collection'' close to your expectations?\\
\-\hspace{0.4cm}- [If No] what is different here/what did not meet your expectations?\\
\-\hspace{0.2cm}- Is the definition of ``data sharing'' close to your expectations?\\
\-\hspace{0.4cm}- [If No] what is different here/what did not meet your expectations?\\
\-\hspace{0.2cm}- Please click on ``Data collection'' under Understand data collection \& data sharing” to view specific exclusions of ``data collection''\\ 
\-\hspace{0.2cm}- Are you surprised about any of the exclusions under ``data collection''?\\
\-\hspace{0.4cm}- Why are you surprised?\\
\-\hspace{0.4cm}- Why are you not surprised\\
\-\hspace{0.2cm}- Please click on ``Data sharing'' under ``Understand data collection \& data sharing'' to view specific exclusions of ``data sharing''\\ 
\-\hspace{0.2cm}- Are you surprised about any of the exclusions under ``data sharing''?\\
\-\hspace{0.4cm}- Why are you surprised?\\
\-\hspace{0.4cm}- Why are you not surprised\\

\subsection{General Feedback}

Feel free to stop screen sharing at this point.\\
\-\hspace{0.2cm}- What do you think of the privacy labels we viewed today?\\
\-\hspace{0.4cm}- Are they something you found useful?\\
\-\hspace{0.6cm}- [if no] why are they not useful?\\
\-\hspace{0.8cm}- [follow-up] What would make them more useful?\\
\-\hspace{0.6cm}- [if yes] What did you find useful?\\
\-\hspace{0.2cm}- Are they something you would find yourself taking a look at in the future when deciding whether or not you want to download an app?\\
\-\hspace{0.4cm}- Why/Why not?\\
\-\hspace{0.2cm}- Out of all the pieces of information we showed you today, which do you think was/will be the most relevant in helping you make a decision if you decide to check it? You don’t have to pick just one piece of information. Tell me about however many you find really important and that you would want to look at.\\
\-\hspace{0.4cm}- Why did you pick these pieces of information?\\
\-\hspace{0.2cm}- Do you have any other feedback or suggestions about the privacy labels we have shown you today?\\

Alright, I have asked all the interview questions. Is there anything else you would like to add/say? Do you have any questions for me at this point? You will receive the bonus through Prolific soon. Feel free to disconnect now. Thank you so much for your participation\! \\

\section{iOS Interview Scripts}
\label{sec:appendix_ios_scripts}

\subsection{Introduction}    

\begin{flushleft}
– Thank you for meeting with me today. This interview is being conducted for research at [ANONYMOUS INSTITUTION] to better understand how people interact with mobile apps in the Apple App Store. I will ask you to answer some questions and view some information in the App Store. This session should take around an hour to complete, and will be recorded via Zoom. You have previously received 40¢ for scheduling the interview, and upon completion of the study, you will be paid \$14.60 through Prolific. You will be asked to share your phone screen via Zoom at some point during the session to enable us to follow what you are doing on your phone as you visit the App store.

– Please answer our questions truthfully and as thoroughly as possible. If in doubt, feel free to ask me for clarification at any point during the interview. I want to emphasize that there are no right or wrong answers. Our goal is simply to understand your opinions and thought processes. You may stop the interview at any point, or choose to not answer a question, or take a break if you wish. Please do not reveal any private or personally-identifiable information about yourself or others during the interview. If you accidentally reveal any personal information, please let me know so that I can remove it from the recording. You have already received the consent form and agreed to it. Do you have any questions at this time? \\

– Please turn off your camera so that your face is not recorded for the interview.\\
\end{flushleft}
\subsection{General Questions about App Usage}
\-\hspace{0.2cm}- When was the last time that you downloaded a new app on your phone? [Followup] Have you downloaded an app in the past year or so?\\
\-\hspace{0.2cm}- What are some of the typical factors that influence which apps you download on your phone? (Prompt: app reviews, brand, ratings, security, ranking, data privacy, your friends) \\
\-\hspace{0.2cm}- Have you ever compared different apps before deciding which one to download (Prompt: what types of things have you compared?)? \\
\-\hspace{0.2cm}- Have you ever considered data privacy in the decisions you make? Why/Why not?\\
\-\hspace{0.2cm}- Have you ever wondered what information/data apps collect about you?\\
\-\hspace{0.4cm}- [If they say they have] How would you go about finding out what information an app collects about you and what the app does with the information?\\
\-\hspace{0.4cm}- [Follow-up] Have you ever actually done that?\\
\-\hspace{0.4cm}- [If they say they have not] If you were to look for this, how could you possibly find out what information an app collects about you and what the app does with that information? Prompt: media? friends/family? experts? privacy policies/EULA? permission settings? Other? \\
\-\hspace{0.4cm}- Do you think those sources are reliable? Have you ever looked?\\
\-\hspace{0.2cm}- Is there any information in the Apple App Store you could turn to if you wanted to consider privacy in your app download decisions? [If they say yes, ask them what sort of information it is and where they would find it.]\\
\-\hspace{0.4cm}- [If they mention the privacy/ data safety label] Do you know what kind of information you would find if you looked at one of these labels?\\
\-\hspace{0.4cm}- [If they don’t mention the privacy/data safety label] Apple has started including app privacy labels for apps in the app store that provide a summary of privacy information since December of 2020. What sort of information would you expect this summary might contain?\\

- [Connect phone to Zoom] If you do not have Zoom on your phone already, please head to the App Store and search for the Zoom app. Download the app on your phone and try to connect to the meeting by clicking on the Zoom link we sent to you earlier. There is no need to join the audio from your phone. There should be a green rectangle on the screen which will allow you to share your phone screen with us. We will be recording what is on your phone screen and thus recommend closing or hiding any apps that you would not like us to accidentally record. You could do so by turning on the ``Do Not Disturb'' feature on your phone. (moon icon) click on start broadcast and open app store \\
At this moment I would like to ask you to turn on ``Do Not Disturb'' from your device. You can do this by swiping down from the top and clicking the circle with a minus sign.\\
\subsection{Just Dance Now}
[App Familiarity] Before we look at any labels,\\  
\-\hspace{0.2cm}- Are you familiar with the Just Dance Now app?\\ 
\-\hspace{0.2cm}- Could you describe what this app does?\\ 
\-\hspace{0.2cm}- Have you used Just Dance Now before? \\ 
\-\hspace{0.4cm}- [If yes] how often do you use it?\\  
\-\hspace{0.4cm}- [If no] It's an app for dancing to music with steps provided by the app on your phone.\\  
We will be taking a look at the privacy label of Just Dance Now. Could you navigate to the App Store and search for the app Just Dance Now? Could you click on the link we sent you. \\ 

Please scroll down to the \textbf{``App Privacy''} section, and now I would like to give you a moment to take a look at the info. \\
\-\hspace{0.2cm}- [overall impression] What is your overall impression of this label? \\
\-\hspace{0.2cm}- Have you seen a label like this before?\\
\-\hspace{0.4cm}- [If yes] Where have you seen it? \\
\-\hspace{0.4cm}- [If yes] What did you do when you saw it? \\
\-\hspace{0.2cm}- [expectations] Is the privacy information you see here in line with what you would expect for an app like this?\\
\-\hspace{0.2cm}- Do you see any information that is surprising or unexpected?
\-\hspace{0.4cm}- [If they did] What surprised you?\\
\-\hspace{0.4cm}- [If they didn't] Why did you expect to see this information?\\
Is there other information about how this app uses data that you would like to know?\\

[Label Comprehension]\\
\-\hspace{0.2cm}- [Tracking] What does the phrase ``data used to track you'' mean to you? \\
\-\hspace{0.4cm}- [Clarification] What do you think an app does with this data?\\
\-\hspace{0.4cm}- [If they say the app will track you] what does tracking mean in this context?\\ 
\-\hspace{0.4cm}- [Follow up if needed with:] How do you think tracking works?
\-\hspace{0.2cm}- Looking at the list of things under ``data used to track you'', does any of this concern you? If so, what? \\
\-\hspace{0.4cm}- Why does this concern you? \\
\-\hspace{0.4cm}- Why does this not concern you? \\
\-\hspace{0.2cm}- [Linked to You] What does the phrase ``data linked to you'' mean to you? \\
\-\hspace{0.4cm}- [Clarification] What do you think an app does with this data?\\
\-\hspace{0.4cm}- [If they say the app will link data to you] what does linking mean in this context? \\
\-\hspace{0.4cm}- How do you think the app does that?\\
\-\hspace{0.2cm}- Looking at the list of things under ``data linked to you'', does any of this concern you? If so, what?\\ 
\-\hspace{0.4cm}- Why does this concern you?\\ 
\-\hspace{0.4cm}- Why does this not concern you?\\
\-\hspace{0.2cm}- [Not Linked to You] Consequently what does the phrase ``data not linked to you'' mean to you? \\
\-\hspace{0.4cm}- [Clarification] What do you think an app does with this data?\\
\-\hspace{0.4cm}- [If they say the app will not link data to you] what does not linking mean in this context?\\
\-\hspace{0.4cm}- How do you think the app does that?\\
\-\hspace{0.4cm}- Looking at the list of things under ``data not linked to you'', does any of this concern you? If so, what? \\
\-\hspace{0.4cm}- Why does this concern you? \\
\-\hspace{0.4cm}- Why does this not concern you?\\

[Record whether the user has clicked on ``See Details'' by this point]\\
\-\hspace{0.2cm}- [if yes] start asking specific questions below\\
\-\hspace{0.2cm}- [If not] ask ``do you see anywhere you can click to get more information about this app’s collection or use of data?'' \\
\-\hspace{0.4cm}- [Note if the user suggests other things they could click on] \\
\-\hspace{0.4cm}- [If they say ``See Details''] let them go ahead and click that\\
\-\hspace{0.4cm}- [if they don’t] suggest ``See Details'' and ask if they can find the ``See Details'' link\\

Take a moment to go over the information. \\
\-\hspace{0.2cm}- [first impression] What is your first impression of the expanded label?\\
\-\hspace{0.2cm}- Have you seen a label expanded like this before? [only if they said they had seen the compact label] \\
\-\hspace{0.4cm}- [If yes] Where have you seen it? \\
\-\hspace{0.4cm}- [If yes] What did you do when you saw it?\\
\-\hspace{0.2cm}- [expectations] Is the privacy information you see here in line with what you would expect for an app like this?\\
Do you see any information that is surprising or unexpected?\\
\-\hspace{0.4cm}- [If they did] What surprised you?\\
\-\hspace{0.4cm}- [If they didn't] Why did you expect to see this information?\\
\-\hspace{0.2cm}- [Only if they mentioned something was missing or needed] Is there anything you mentioned you’d like to see in the compact label present in the expanded label? \\
\-\hspace{0.4cm}- Is there anything here that surprises you?\\
\-\hspace{0.4cm}- Is there anything here that you don’t understand?\\
Let’s look specifically at \textbf{Data Used to Track You}. \\
\-\hspace{0.2cm}- Do you see any details here that you didn’t see before in the compact label? If so, what?\\
\-\hspace{0.2cm}- Is there anything here that surprises you?\\
\-\hspace{0.2cm}- Is there anything here that you don’t understand? \\
Now let’s look at \textbf{Data Linked to You}.\\
\-\hspace{0.2cm}- Do you see any details here that you didn’t see before in the compact label? If so, what?\\
\-\hspace{0.2cm}- There are three groups here: Analytics,  Product Personalization, and App functionality. Why do you think there are three groups?\\
\-\hspace{0.2cm}- Is there anything here that surprises you?\\
\-\hspace{0.2cm}- Is there anything here that you don’t understand?\\ 
Now let’s take a look at \textbf{Data Not Linked to You}.\\
\-\hspace{0.2cm}- What do you see under this heading?\\
\-\hspace{0.2cm}- There is one group here: App Functionality. Why do you think there is only one group here?\\
\-\hspace{0.2cm}- Is there anything here that surprises you?\\
\-\hspace{0.2cm}- Is there anything here that you don’t understand?\\ 

Looking over this entire expanded label, what do you see here that you care about most? \\
\-\hspace{0.2cm}- Why do you care about it?\\ 
Is there anything in the expanded label that changed how concerned or not concerned you are about this app?\\
\-\hspace{0.2cm}- Who do you think created this label?\\
\-\hspace{0.4cm}- [follow-up] Which company do those people work for?\\
How confident are you that this label accurately describes how this app will use your data?\\
Is there any other information not covered here that you would still like to know?\\
\-\hspace{0.2cm}- [If yes] where would you go to find it?\\
\-\hspace{0.2cm}- [If no] Suppose you had additional questions about this information, where would you go to find it?\\
\-\hspace{0.4cm}- [If privacy policy mentioned] Where would you find this policy?  (developer’s privacy policy)\\

\subsection{Google Maps}

\-\hspace{0.2cm}- Now we are going to read the privacy label of a different app. Are you familiar with Google Maps? \\
\-\hspace{0.2cm}- Could you describe what this app does? Google Maps is a web-based service that provides detailed information about geographical regions and sites around the world.\\
\-\hspace{0.2cm}- Have you used Google Maps before? \\
\-\hspace{0.4cm}- [If yes] how often do you use it?\\
\-\hspace{0.4cm}- [if not, explain:] \\

Ok, let’s now take a look at Google Maps’s privacy label. Could you search for Google Maps in the App Store? Please scroll down to the ``App Privacy'' section. \\
\-\hspace{0.2cm}- [overall impression] What is your overall impression of this label? \\
\-\hspace{0.4cm}- Do any differences between this label and the previous label jump out?\\
\-\hspace{0.2cm}- [expectations] Is the privacy information you see here in line with what you would expect for an app like this?\\
\-\hspace{0.2cm}- Do you see any information that is surprising or unexpected?\\
\-\hspace{0.4cm}- [If they did] What surprised you?\\
\-\hspace{0.4cm}- [If they didn't] Why did you expect to see this information?\\
\-\hspace{0.2cm}- Is there other information about how this app uses data that you would like to know?\\

[Label Comprehension]
\-\hspace{0.2cm}- (optional, probably skip) What do you notice about this privacy label compared to the label we just saw for Just Dance Now? What is not present here? How is Google Maps’s compact label different from Just Dance Now's compact label? What is your overall impression of this? [prompt user to talk about how there’s only Data Linked to You” for Google Maps]\\

\-\hspace{0.2cm}- Do you think Google Maps is using your data to track you? 
\-\hspace{0.4cm}- [if answered yes, ask why they think that]  \\
\-\hspace{0.4cm}- [if misunderstand tracking, clarify def and re-ask] ``Tracking'' refers to linking data they collect with other company’s data in order to select ads targeted at you \\

\-\hspace{0.2cm}- Do you think Google Maps collects any data not linked to you? [if answered yes, ask why they think that]\\

\-\hspace{0.2cm}- How secure do you think Google Maps is?\\
\-\hspace{0.4cm}- Why do you think it is secure?\\
\-\hspace{0.4cm}- Why do you think it is not secure?\\
\-\hspace{0.2cm}- What would make you more confident about the security of Google Maps?\\
\-\hspace{0.4cm}- [follow-up] would you be more confident if Google Maps had passed an independent security review?\\ 

Please click on \textbf{``See details''} to expand the privacy label. \\
\-\hspace{0.2cm}- Are you surprised about anything you see in the expanded label?\\ 
\-\hspace{0.4cm}- [If so] Which information do you find surprising to see here?\\
\-\hspace{0.4cm}- [If not, don't mention third-party ads] Why are you not surprised about data being used for third-party advertising? \\
\-\hspace{0.4cm}- [follow-up] In that case, why do you think this app does not have any data used to track you?\\

\subsection{Tumblr - Fandom, Art, Chaos}

\-\hspace{0.2cm}- Now we are going to read the privacy label of a different app. \\
\-\hspace{0.2cm}- Are you familiar with Tumblr? \\
\-\hspace{0.2cm}- Could you describe what this app does? [Tumblr is a microblogging and social networking platform where users could post multimedia and other content to a short-form blog]\\
\-\hspace{0.2cm}- Have you used Tumblr before? \\
\-\hspace{0.4cm}- [If yes] how often do you use it?\\
\-\hspace{0.4cm}- [if not, explain:] \\

Could you please exit the current label and find Tumblr – Fandom, Art, Chaos” in the App Store? Please scroll down to the ``App Privacy'' section. \\
\-\hspace{0.2cm}- [overall impression] What is your overall impression of this label? \\
\-\hspace{0.4cm}- Do any differences between this label and the previous labels jump out?\\
\-\hspace{0.2cm}- [expectations] Is the privacy information you see here in line with what you would expect for an app like this?\\
\-\hspace{0.2cm}- Do you see any information that is surprising or unexpected?
\-\hspace{0.4cm}- [If they did] What surprised you?\\
\-\hspace{0.4cm}- [If they didn't] Why did you expect to see this information?\\
\-\hspace{0.2cm}- Is there other information about how this app uses data that you would like to know?\\

[Note whether the user has opened ``See Detail'' by now, if not, ask them where they could find more privacy information, ask them to click on ``See Details'' if they could not find it]\\
\-\hspace{0.2cm}- How securely do you think Tumblr handles your data?\\
\-\hspace{0.2cm}- [Data Encryption] Have you heard of ``data encryption''?\\
\-\hspace{0.4cm}- [If so], what does it mean? \\
\-\hspace{0.4cm}- [If not] data encryption means that information is converted into secret code that hides the information and helps make sure that nobody can eavesdrop and see the information. Some apps use data encryption whenever they send information to the server so that eavesdroppers can’t read your personal information. \\
\-\hspace{0.2cm}- Do you think that Tumblr encrypts messages it sends?\\
\-\hspace{0.2cm}- What do you think will happen if your data isn’t encrypted in transit?\\
\-\hspace{0.2cm}- Would you like the privacy label to tell you whether or not an app uses data encryption? \\
\-\hspace{0.2cm}- If Tumblr did not use encryption, would you be less likely to download it?\\
\-\hspace{0.4cm}- Why / Why Why not? \\
\-\hspace{0.2cm}- Who do you think could get access to the unencrypted data?\\
\-\hspace{0.2cm}- [Data Deletion] Do you think that Tumblr has a feature that allows you to delete all the data it stores about you? \\
\-\hspace{0.2cm}- Are apps required to delete your data if you request that data be deleted? \\
\-\hspace{0.2cm}- What do you think happens to the data after you request for it to be deleted?\\
\-\hspace{0.2cm}- Would you like the privacy label to tell you whether or not you could request for your data to be deleted? \\
\-\hspace{0.2cm}- If Tumblr did not allow you to request to delete your data, would you be less likely to download it?\\

[Advertising/Marketing] 

Let’s take a look at ``Data Linked to You''\\
\-\hspace{0.2cm}- Do you see anything here that is related to advertising or marketing?\\
\-\hspace{0.2cm}- What does ``third-party advertising'' mean to you?\\
\-\hspace{0.2cm}- What does ``developer’s advertising or marketing'' mean to you?\\
\-\hspace{0.4cm}- [if not obviously different] What is the difference between data used for ``third-party advertising'' and data used for ``developer’s advertising or marketing''? \\
\-\hspace{0.2cm}- [follow-up] Who is doing the advertising here?  \\
\-\hspace{0.2cm}- [follow up] what exactly is a third party or developer? [if they mention these in their answers above]\\
\-\hspace{0.2cm}- [Location] Do you think this app is using your location to personalize content/ads?\\

\subsection{Reading Definitions}

\-\hspace{0.2cm}- If you want to learn more about what ``Tracking'' and ``Data Linking'' mean, where would you go to find this information?\\
\-\hspace{0.4cm}- [If they missed it] Tell them to go to the expanded label and click on ``Privacy Definitions and Examples'' and \textbf{keep scrolling down until you see ``Data linked to you''} \\
\-\hspace{0.2cm}- Take a moment and go over the definitions and please let me know when you have finished reading them.\\
\-\hspace{0.2cm}- Is the definition of ``Data linked to you'' close to what you would expect?\\
\-\hspace{0.4cm}- [if not] what is different here/ did not meet your expectations?\\
\-\hspace{0.2cm}- Is the definition of ``Data not linked to you'' close to what you would expect?\\
\-\hspace{0.4cm}- [if not] what is different here/ did not meet your expectations?\\
\-\hspace{0.2cm}- Is the definition of ``tracking'' close to what you would expect?\\
\-\hspace{0.4cm}- [If not] what is different here/ did not meet your expectations?\\
\-\hspace{0.2cm}- Is the definition of ``third-party'' close to what you would expect?\\
\-\hspace{0.4cm}- [If not] what is different here/ did not meet your expectations?\\

\subsection{General Feedback}

Feel free to stop screen sharing at this point.\\
\-\hspace{0.2cm}- What do you think of the privacy labels we viewed today?\\
\-\hspace{0.4cm}- Are they something you found useful?\\
\-\hspace{0.6cm}- [if not] why are they not useful?\\
\-\hspace{0.8cm}- [follow-up] What would make them more useful?\\
\-\hspace{0.6cm}- [if yes] What did you find useful?\\
\-\hspace{0.4cm}- Are they something you would find yourself taking a look at in the future when deciding whether or not you want to download an app?\\
\-\hspace{0.6cm}- Why/Why not?\\
\-\hspace{0.4cm}- Out of all the pieces of information we showed you today, which do you think was/will be the most relevant in helping you make a decision if you decide to check it? You don’t have to pick just one piece of information. Tell me about however many you find really important and that you would want to look at.
\-\hspace{0.6cm}- Why did you pick these pieces of information?\\
\-\hspace{0.4cm}- Do you have any other feedback or suggestions about the privacy labels we have shown you today?\\

I have asked all the interview questions. Is there anything else you would like to add/say? Do you have any questions for me at this point? You will receive the bonus through Prolific soon. Feel free to disconnect now. Thank you so much for your participation! \\

\section{Data Collection and Sharing Exclusions on Google Play}
\label{sec:appendix_google_exclusions}

This information was extracted from Google's online documentation at https://support.google.com/googleplay/answer/11416267.

\subsection{Data Collection Exclusions}
\label{sec:collection_exclusions}

Developers do not need to disclose data accessed by an app as ``collected'' in the Data safety section if:

\begin{itemize}
    \item \textbf{An app accesses the data only on your device and it is not sent off your device.} For example, if you provide an app permission to access your location, but it only uses that data to provide app functionality on your device and does not send it to its server, it does not need to disclose that data as collected.
    \item \textbf{Your data is sent off the device but only processed eph-emerally.} This means the developer accesses and uses your data only when it is stored in memory, and retains the data for no longer than necessary to service a specific request. For example, if a weather app sends your location off your device to get the current weather at your location, but the app only uses your location data in memory and does not store the data for longer than necessary to provide the weather.
    \item \textbf{Your data is sent using end-to-end encryption.} This means the data is unreadable by anyone other than the sender and recipient. For example, if you send a message to a friend using a messaging app with end-to-end encryption, only you and your friend can read the message.
\end{itemize}

\subsection{Data Sharing Exclusions}
\label{sec:sharing_exclusions}
In some cases, app developers do not need to declare data that is transferred to others as ``shared'' in the Data safety section. This includes when:

\begin{itemize}
    \item The data is transferred to a third party based on a specific action that you initiate, where you reasonably expect the data to be shared. For example, when you send an email to or share a document with another person.
    \item The data transfer to a third party is prominently disclosed in the app, and the app requests your consent in a way that meets the requirements of Google Play's User Data policy.
    \item The data is transferred to a service provider to process it on the developer's behalf. For example, a developer may use a service provider to host data on their behalf and in compliance with the developer's instructions, contractual terms, privacy policies, and security standards.
    \item The data is transferred for specific legal purposes, such as in response to a government request.
    \item The data transferred is fully anonymized so it can no longer be associated with any individual.

\end{itemize}


\section{Android and iOS Privacy Labels}\label{appendix:label_screenshots}
In this section, we present the complete Android and iOS labels that were shown to the participants. The Android and iOS labels of Just Dance Now and Google Maps, as well as the iOS label of Tumblr, were taken directly from the respective app webpages. The Android label of Tumblr was extracted from an interview video recording due to the label update that occurred after our interview. The screenshots displayed here match the ones used during the interviews.
\begin{figure}[ht!]
  \begin{framed}
    \includegraphics[width=.9\textwidth, bb=0 0 562 408]{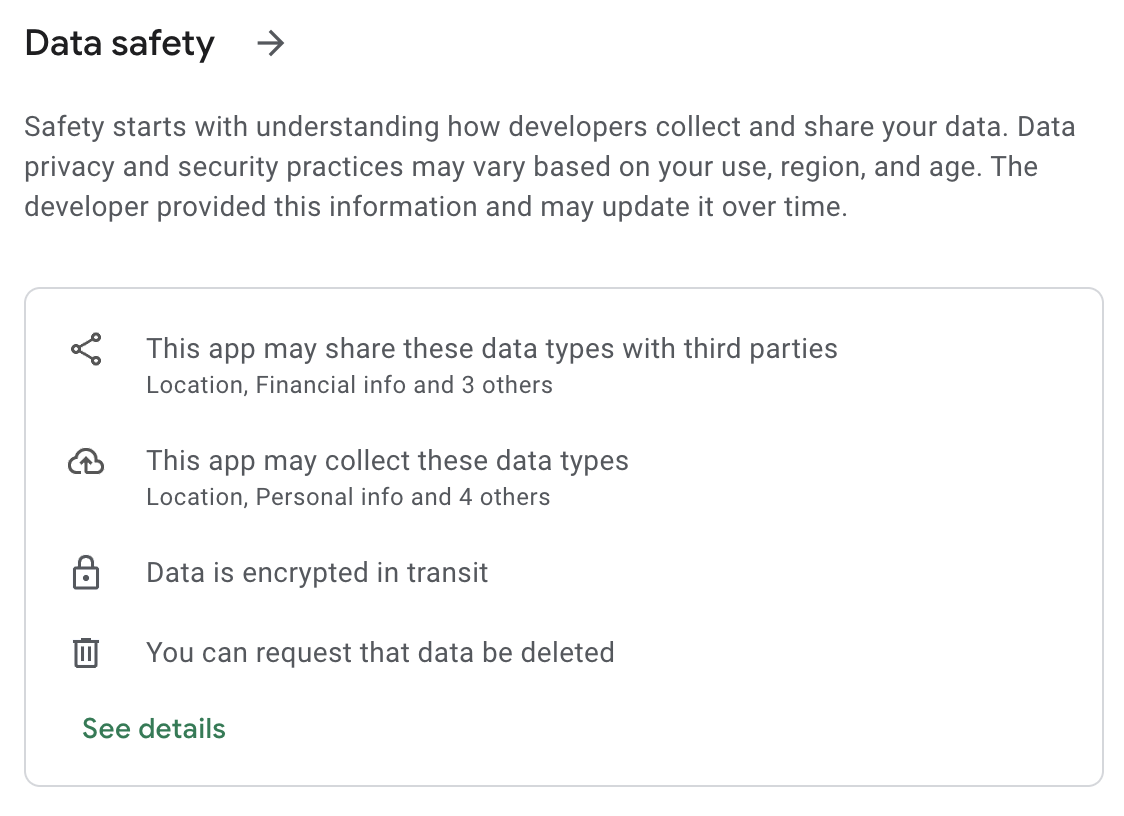}
  \end{framed}
  \caption{Compact Privacy Label of Just Dance Now (Android)}
\end{figure}

\begin{figure}[ht!]
    \begin{framed}
        \includegraphics[width=.9\textwidth]{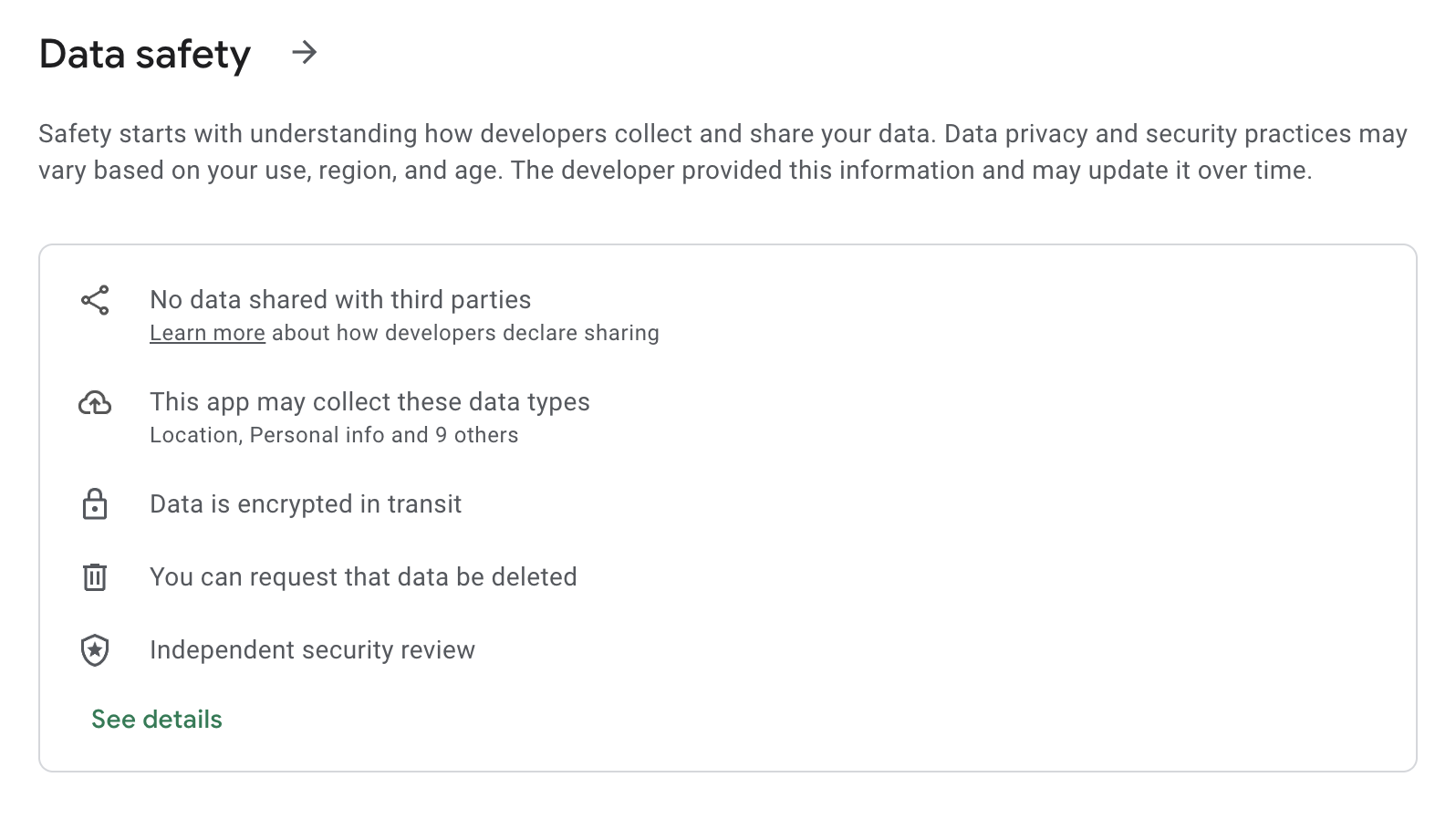}
    \end{framed}
     \caption{Compact Privacy Label of Google Maps (Android)}
\end{figure}

\begin{figure}[ht!]
    \begin{framed}
        \includegraphics[width=.7\textwidth]{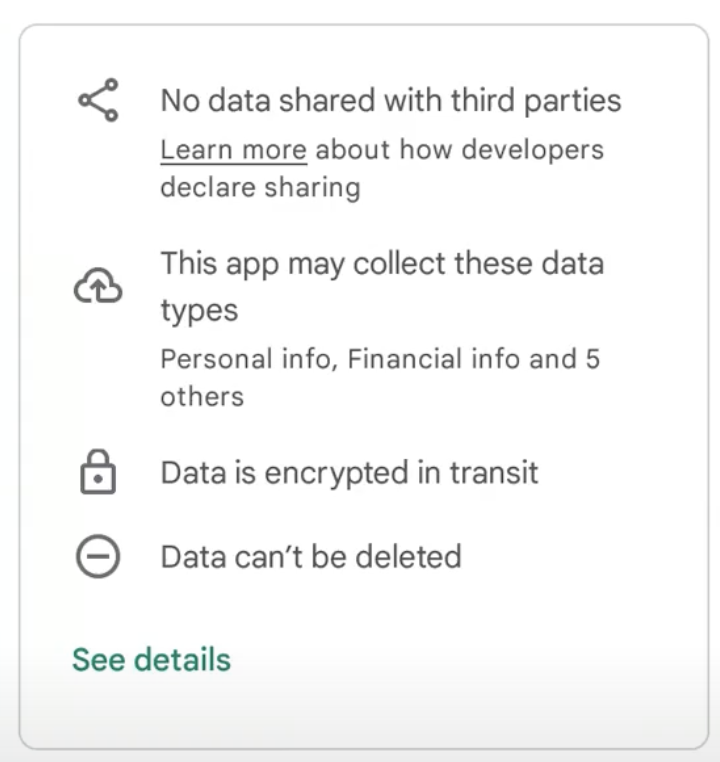}
    \end{framed}
    \caption{Compact Privacy Label of Tumblr-Fandom, Art, Chaos (Android)}
        \begin{flushleft}
            This image was captured from a phone recording to reflect Tumblr's label at the time of the interview.
        \end{flushleft}
\end{figure}

\begin{figure}[ht!]
    \begin{framed}
        \includegraphics[width=\textwidth]{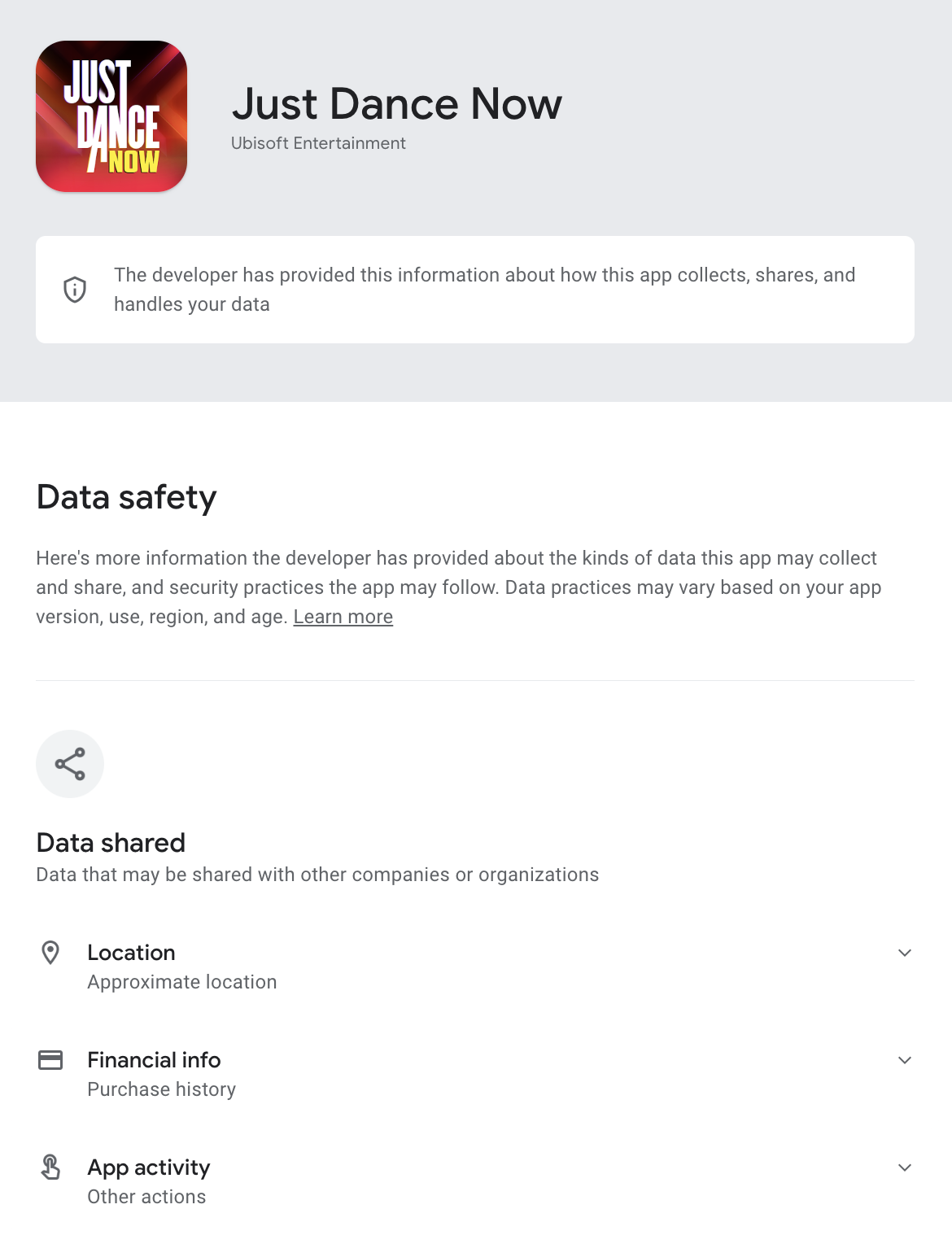}
        \includegraphics[width=\textwidth]{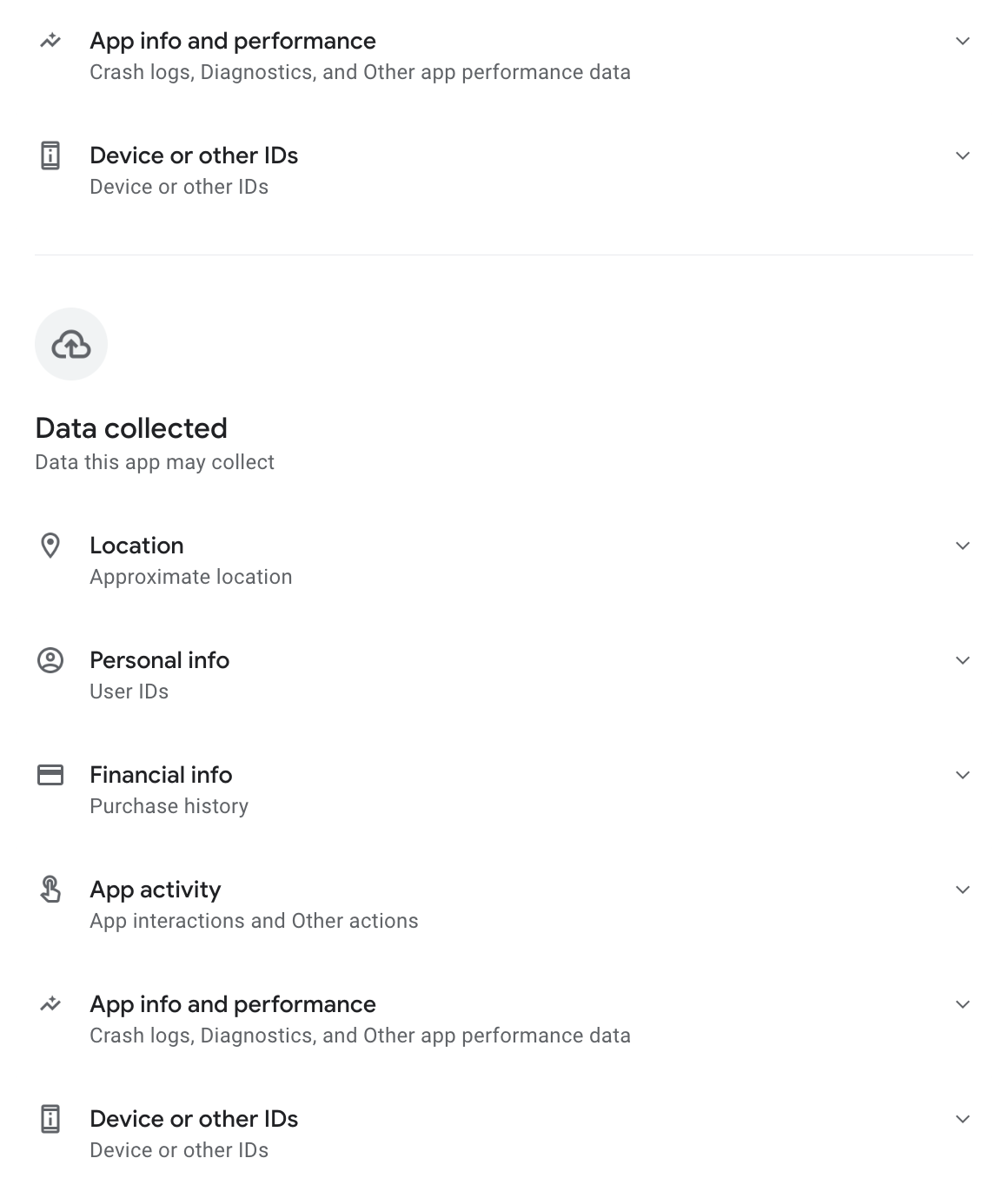}
    \end{framed}
    \caption{Expanded Privacy Label of Just Dance Now Part 1 (Android)}
\end{figure}

\begin{figure}[ht!]
    \begin{framed}
       \includegraphics[width=\textwidth]{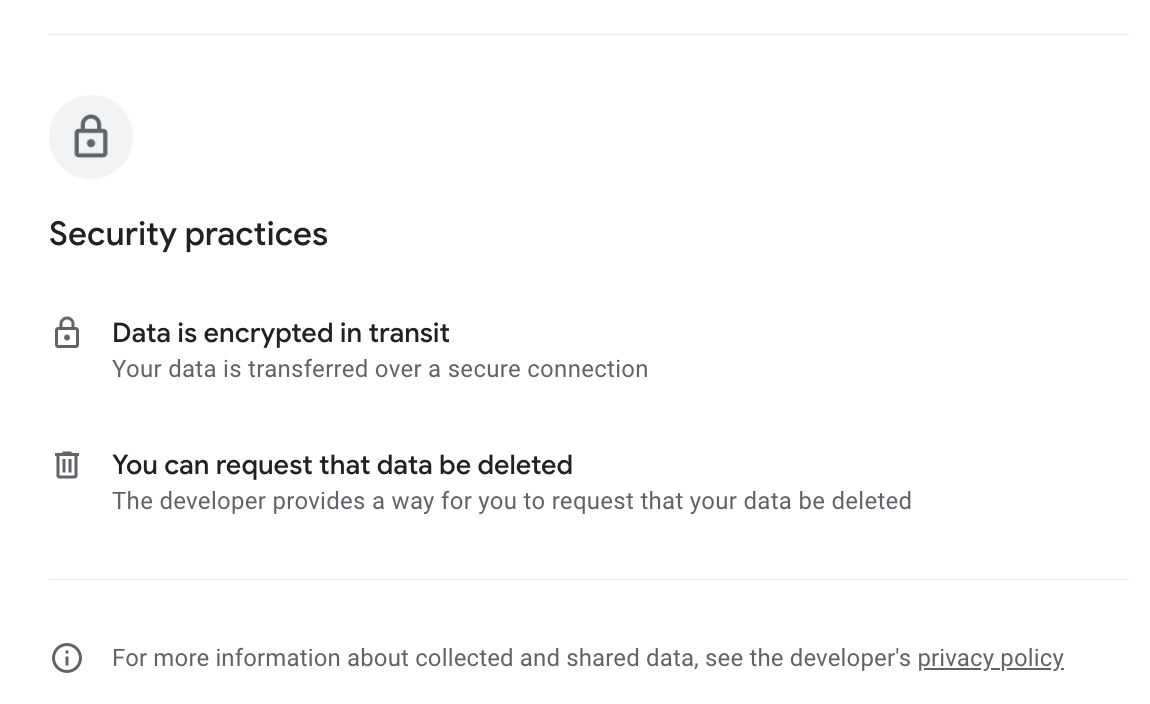}
    \end{framed}
    \caption{Expanded Privacy Label of Just Dance Now Part 2 (Android)}
\end{figure}

\begin{figure}[ht!]
    \begin{framed}
        \includegraphics[width=\textwidth]{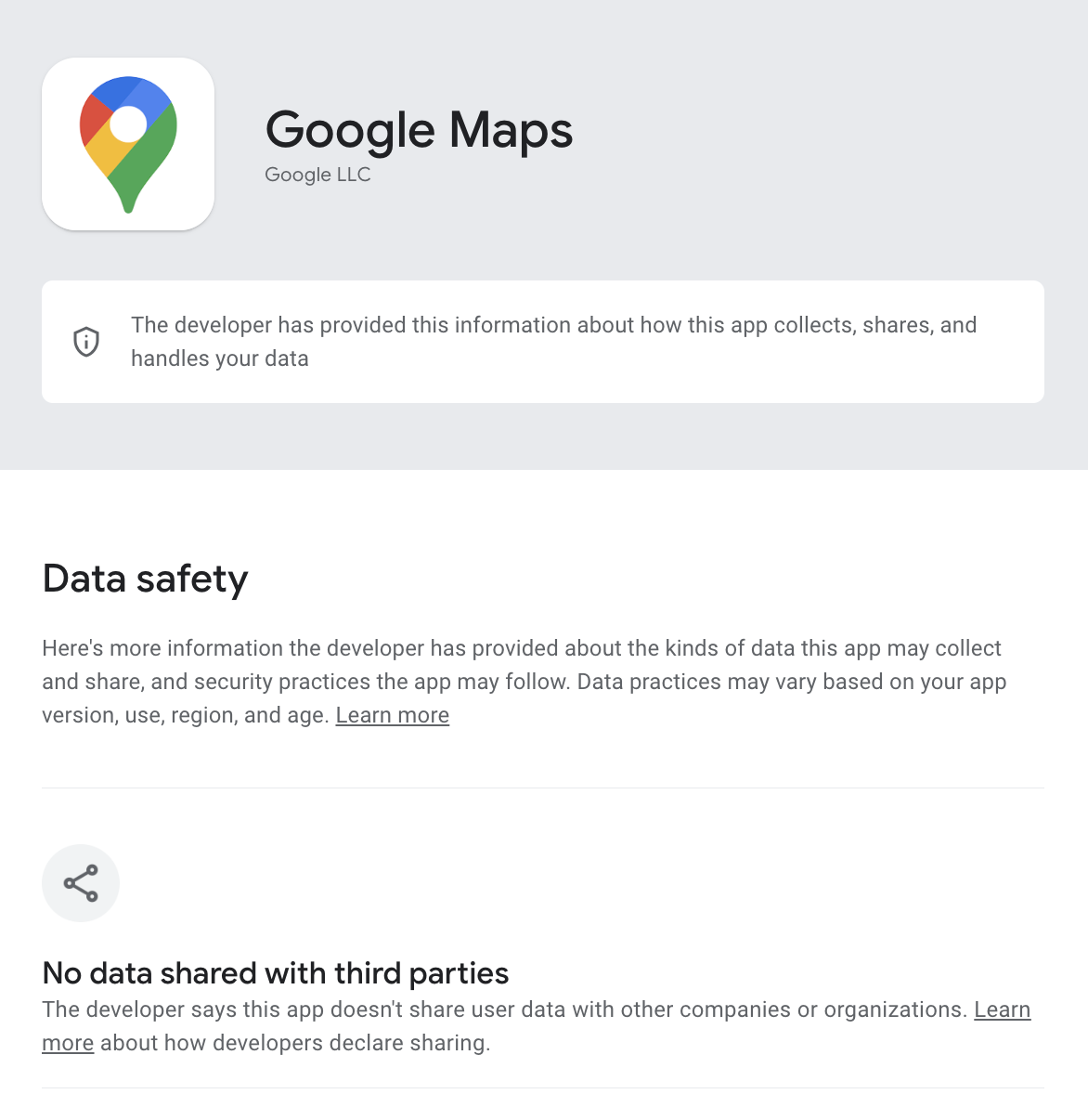}
        \includegraphics[width=\textwidth]{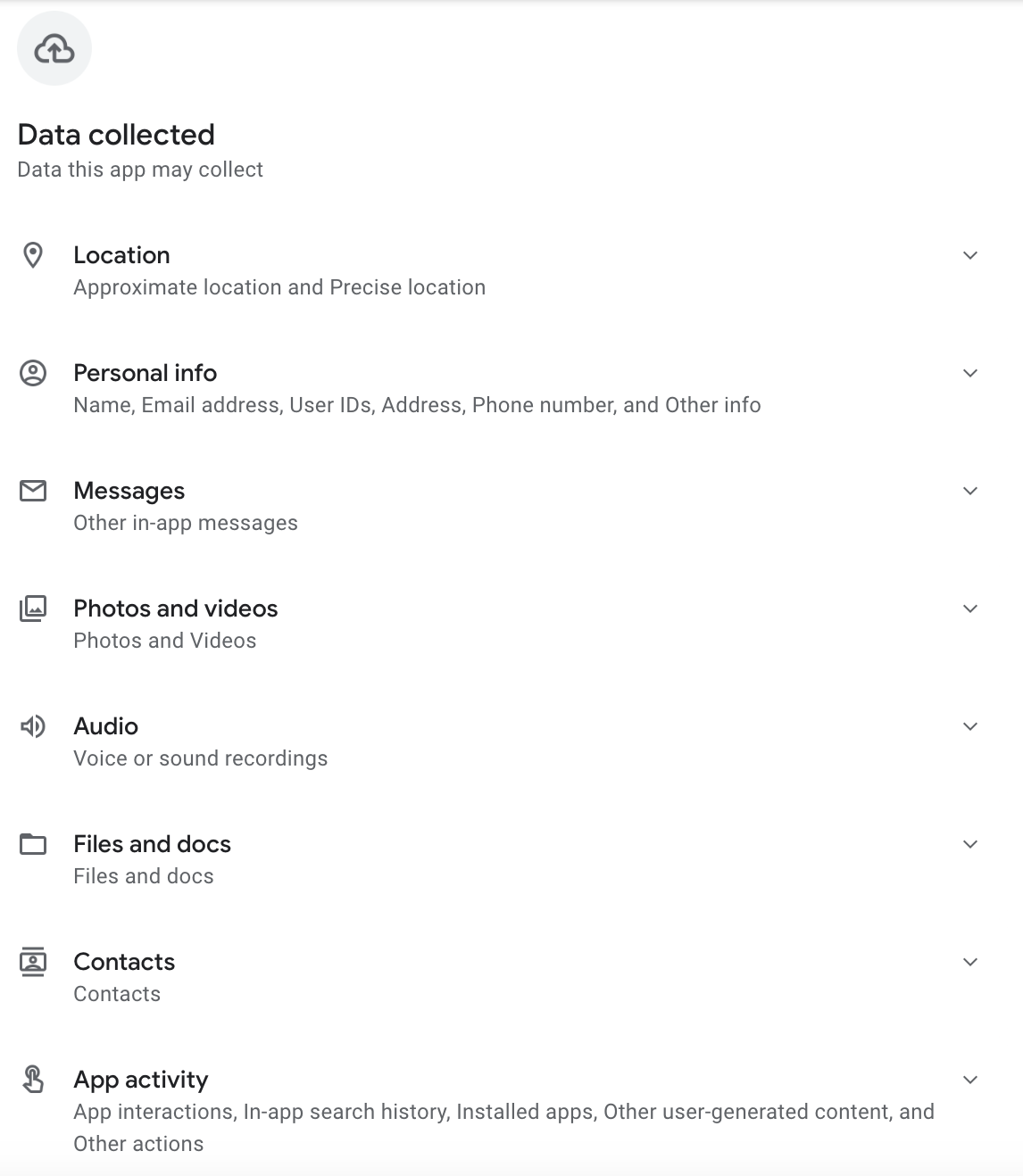}
    \end{framed}
    \caption{Expanded Privacy Label of Google Maps Part 1 (Android)}
\end{figure}

\begin{figure}[ht!]
    \begin{framed}
        \includegraphics[width=\textwidth]{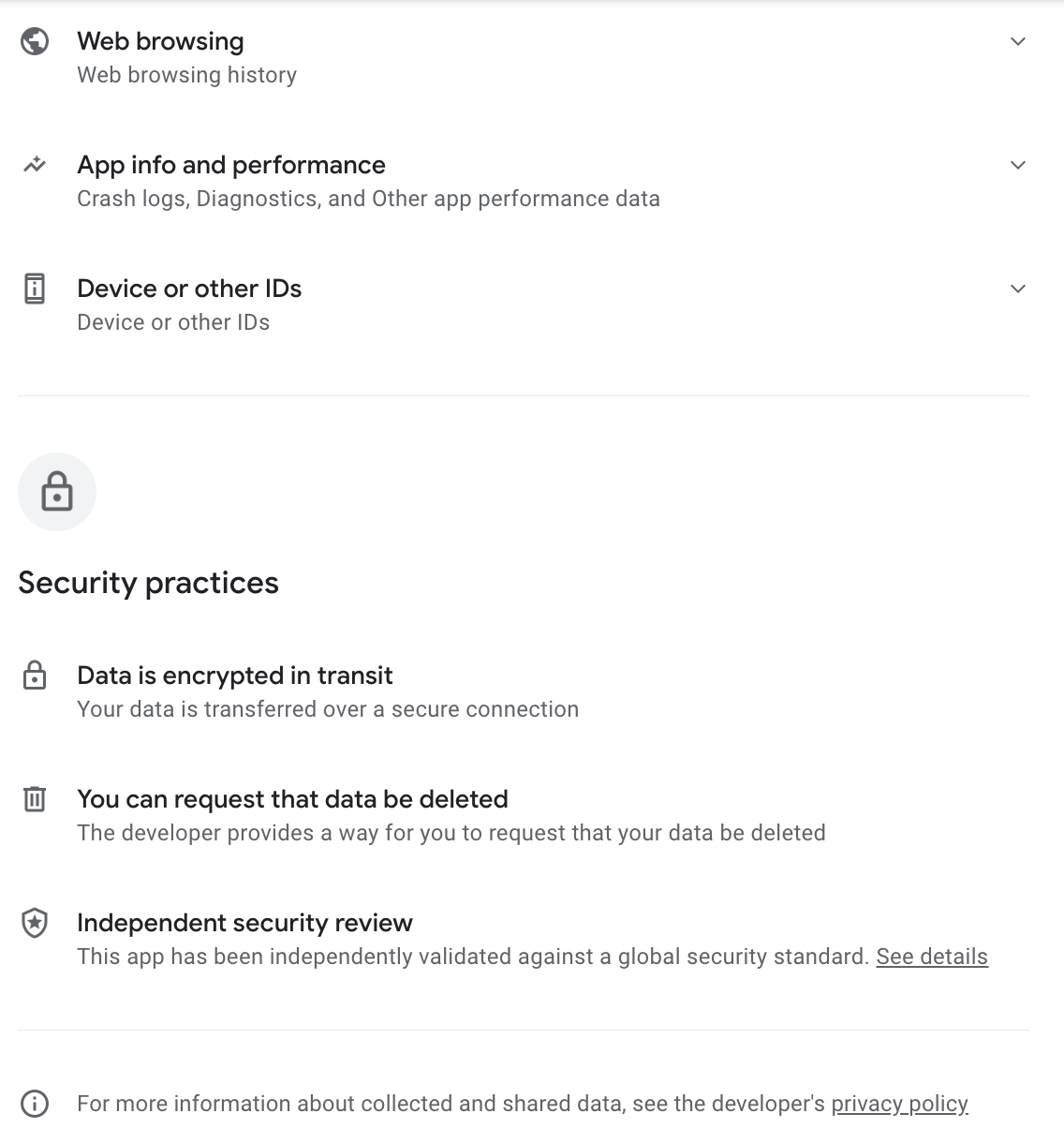}
    \end{framed}
    \caption{Expanded Privacy Label of Google Maps Part 2 (Android)}
\end{figure}

\begin{figure}[ht!]
    \begin{framed}
        \includegraphics[width=\textwidth]{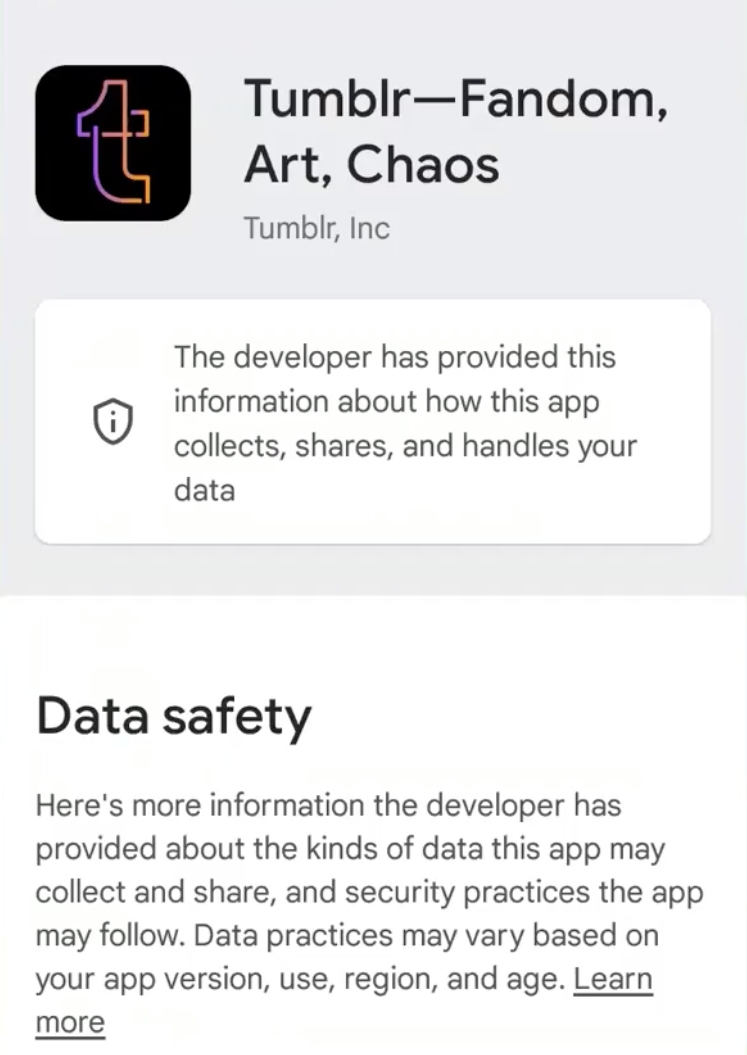}
        \includegraphics[width=\textwidth]{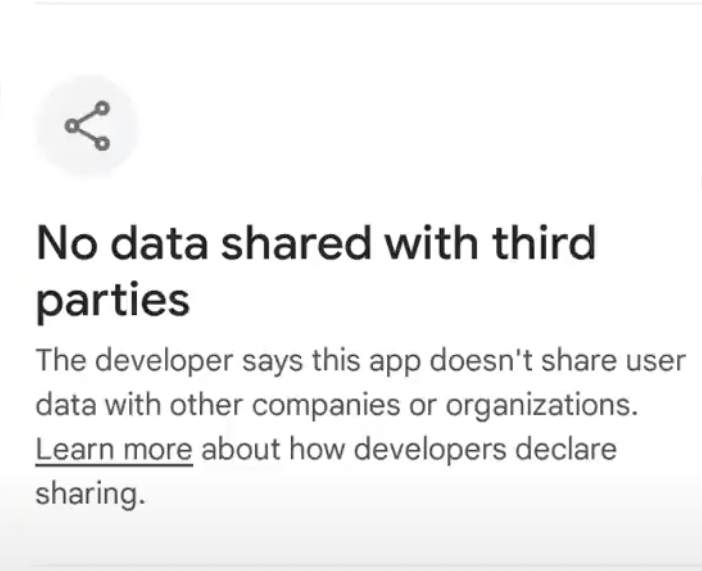}
    \end{framed}
    \caption{Expanded Privacy Label of Tumblr Part 1 (Android)}
        \begin{flushleft}
            This image was captured from a phone recording to reflect Tumblr's label at the time of the interview.
        \end{flushleft}
\end{figure}

\begin{figure}[ht!]
    \begin{framed}
        \includegraphics[width=.9\textwidth]{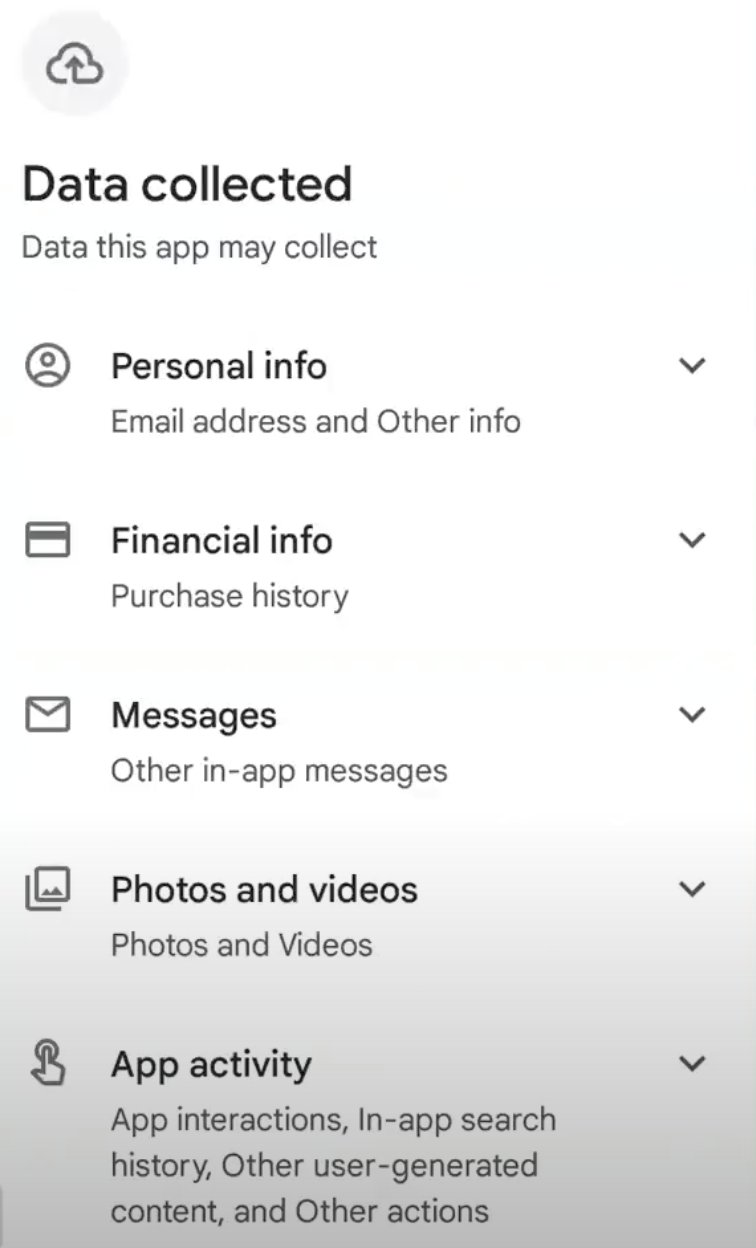}
        \includegraphics[width=.9\textwidth]{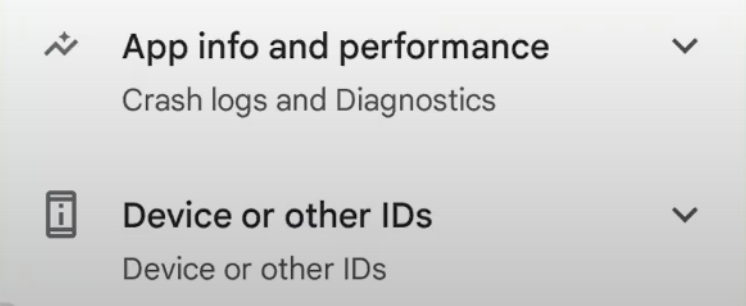}
        \includegraphics[width=\textwidth]{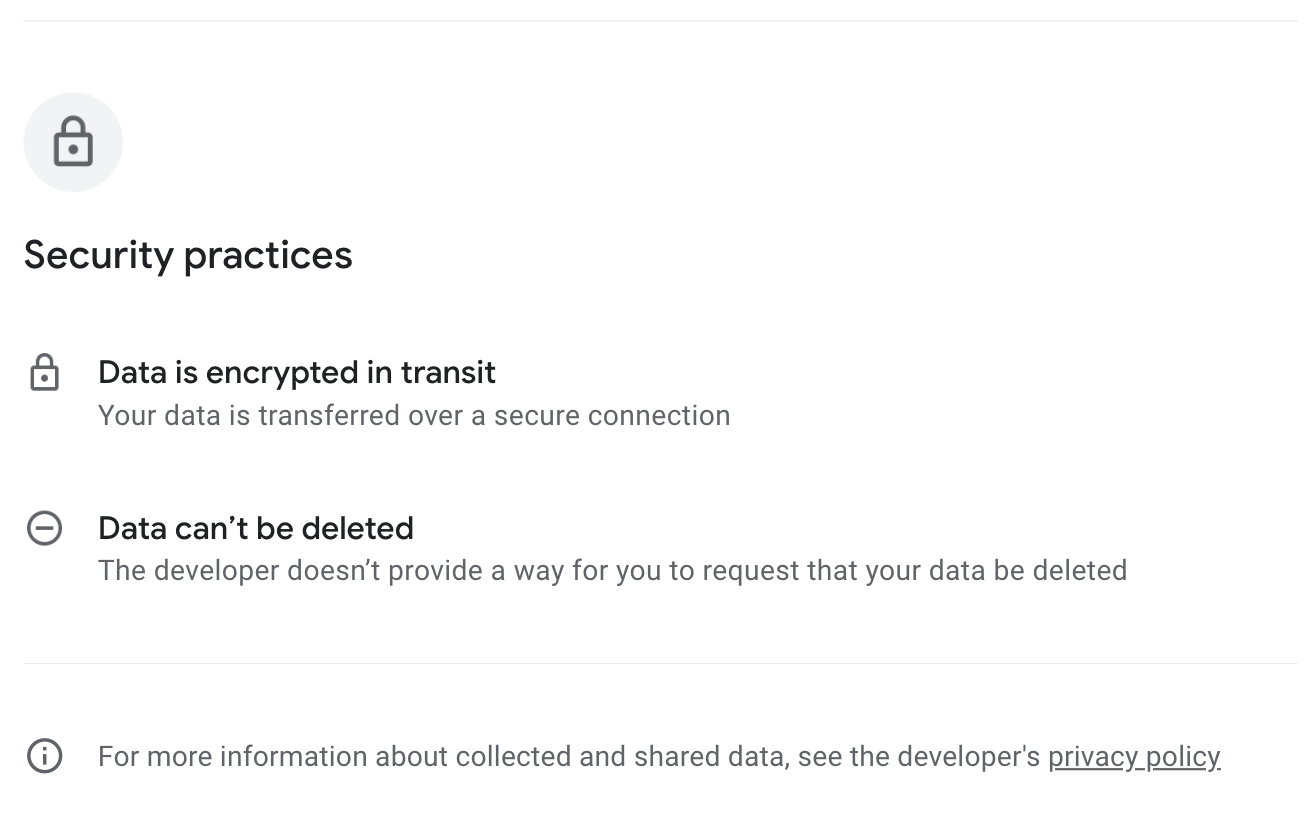}
    \end{framed}
    \caption{Expanded Privacy Label of Tumblr Part 2 (Android)}
        \begin{flushleft}
            This image was captured from a phone recording to reflect Tumblr's label at the time of the interview.
        \end{flushleft}
\end{figure}

\begin{figure}[ht!]
    \begin{framed}
        \includegraphics[width=\textwidth]{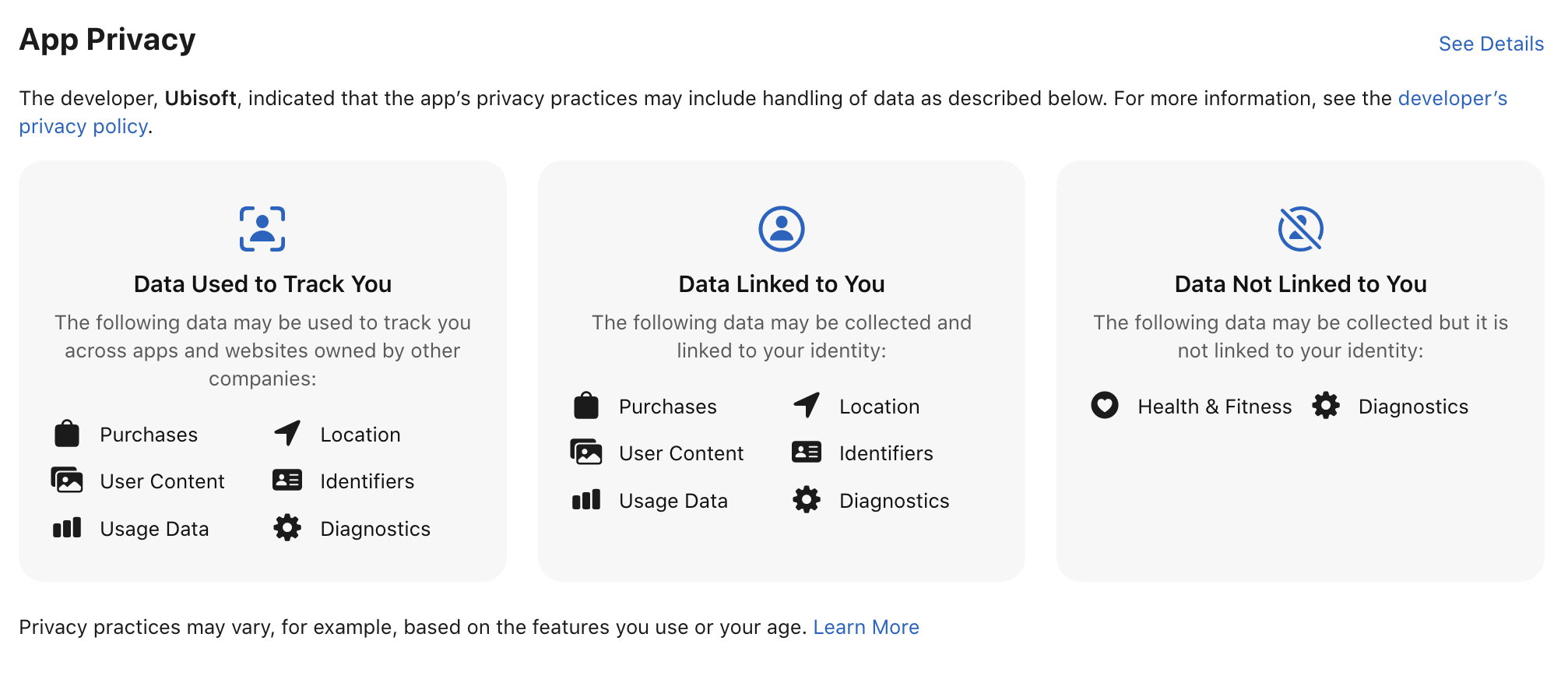}
    \end{framed}
    \caption{Compact Privacy Label of Just Dance Now (iOS)}
\end{figure}

\begin{figure}[ht!]
    \begin{framed}
        \includegraphics[width=\textwidth]{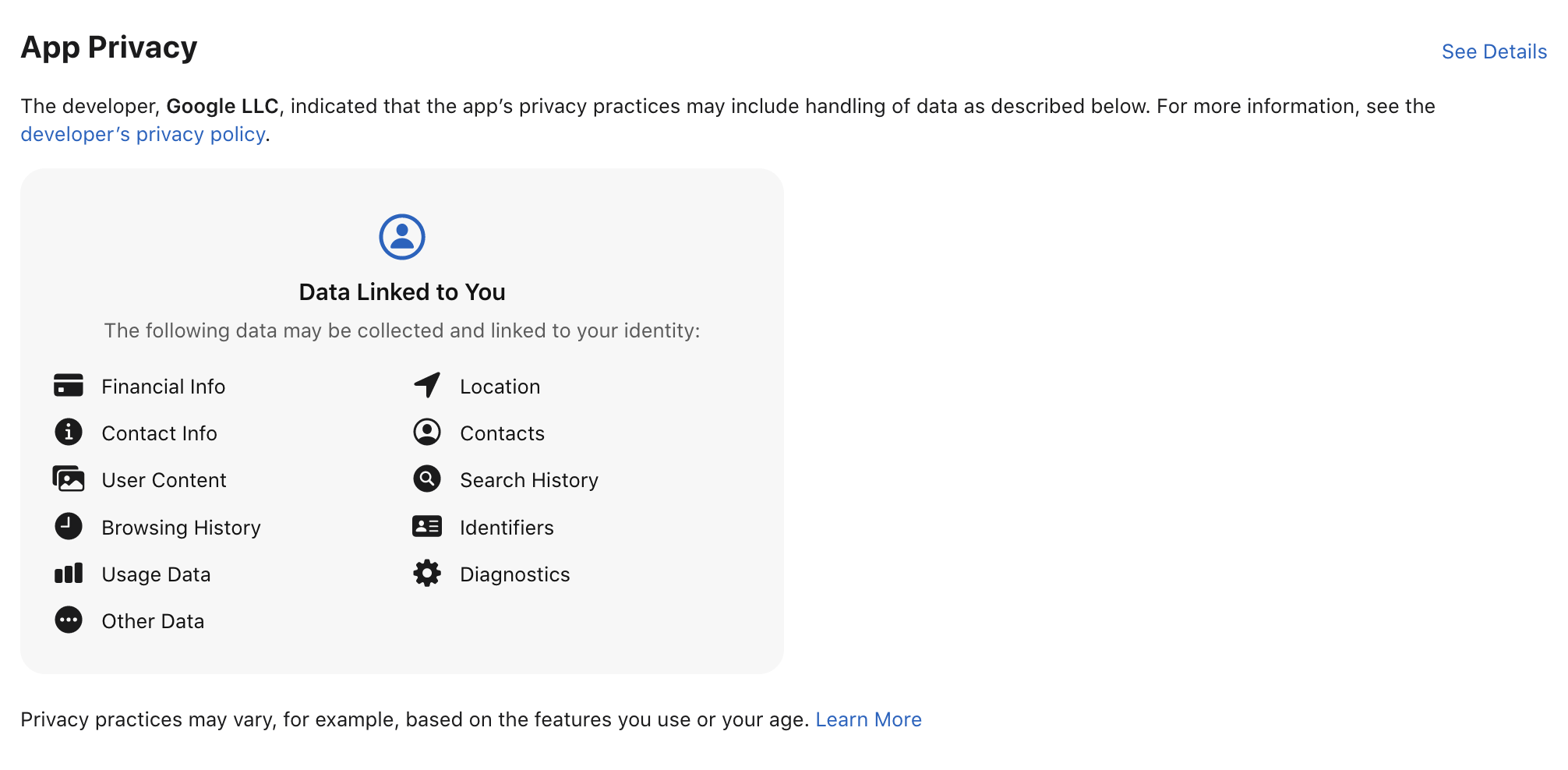}
    \end{framed}
    \caption{Compact Privacy Label of Google Maps (iOS)}
\end{figure}

\begin{figure}[ht!]
    \begin{framed}
    \includegraphics[width=\textwidth]{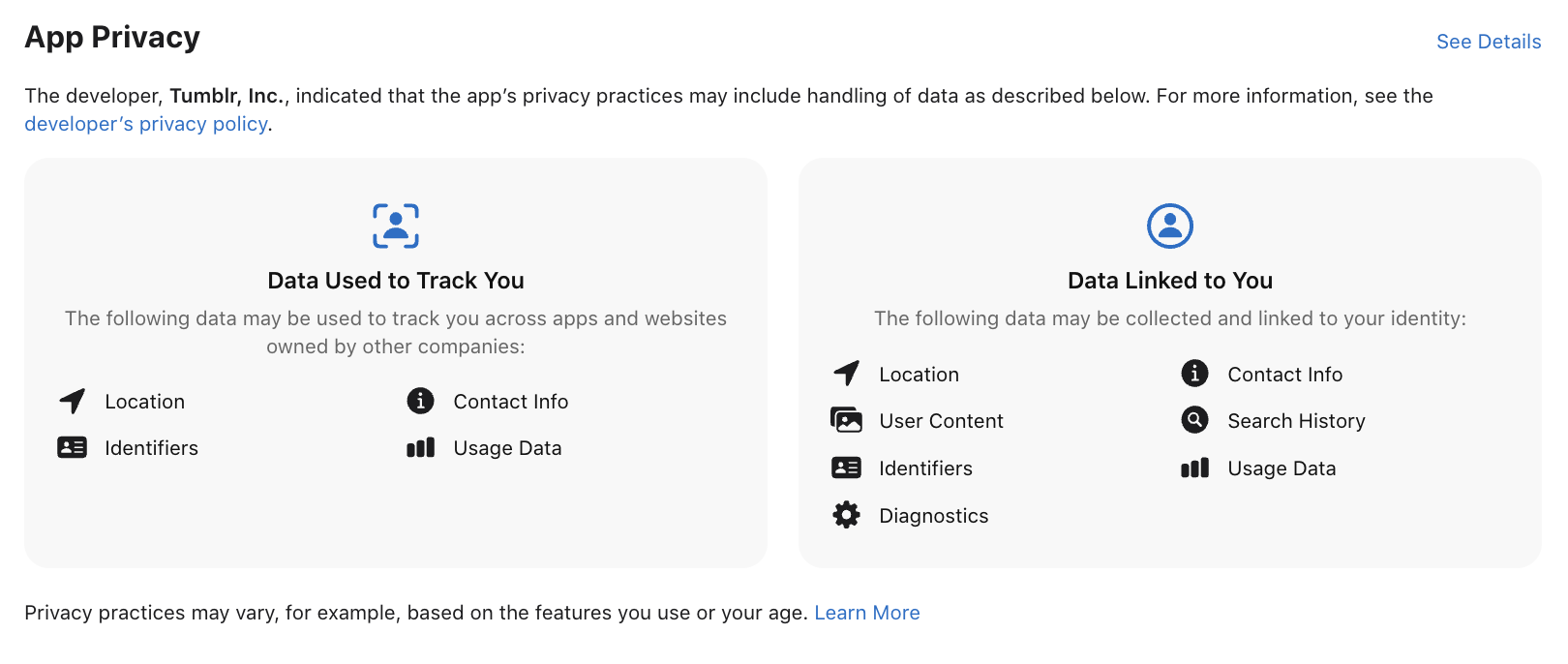}
    \end{framed}
    \caption{Compact Privacy Label of Tumblr-Fandom, Art, Chaos (iOS)}
\end{figure}

\begin{figure}[ht!]
    \begin{framed}
        \includegraphics[width=\textwidth]{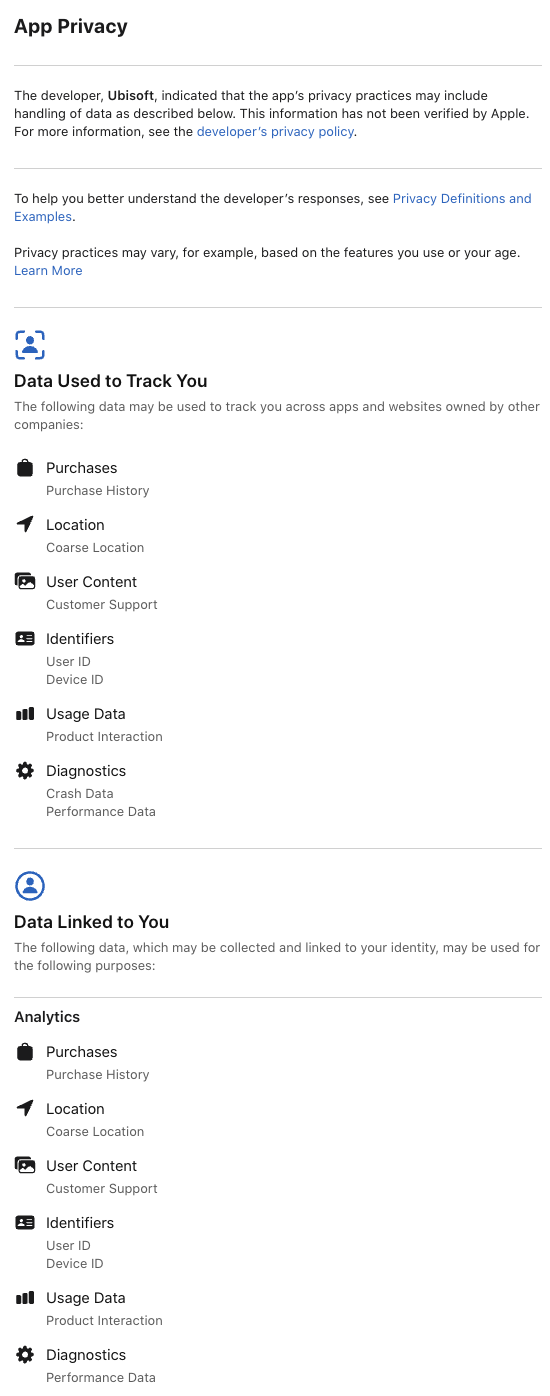}
    \end{framed}
    \caption{Expanded Privacy Label of Just Dance Now Part 1 (iOS)}
\end{figure}

\begin{figure}[ht!]
    \begin{framed}
        \includegraphics[width=\textwidth]{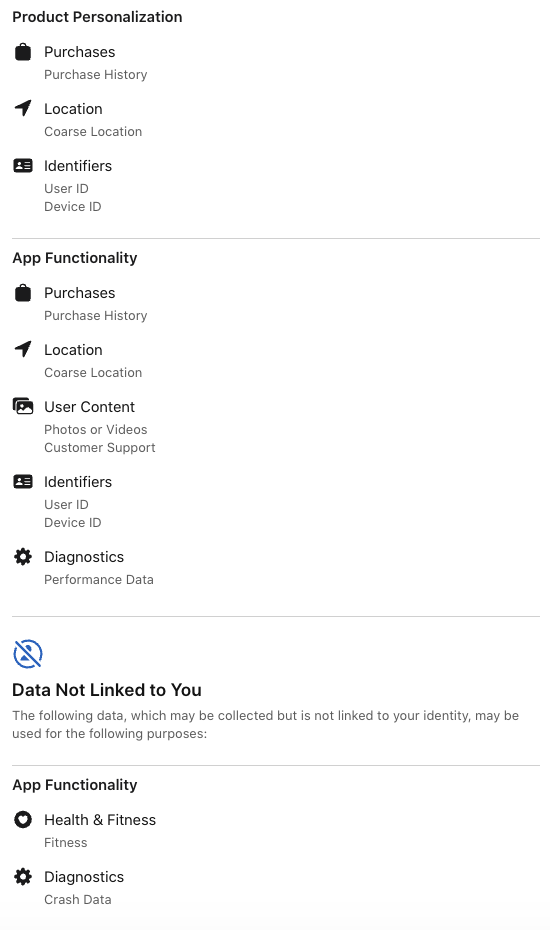}
    \end{framed}
    \caption{Expanded Privacy Label of Just Dance Now Part 2 (iOS)}
\end{figure}

\begin{figure}[ht!]
    \begin{framed}
        \includegraphics[width=\textwidth]{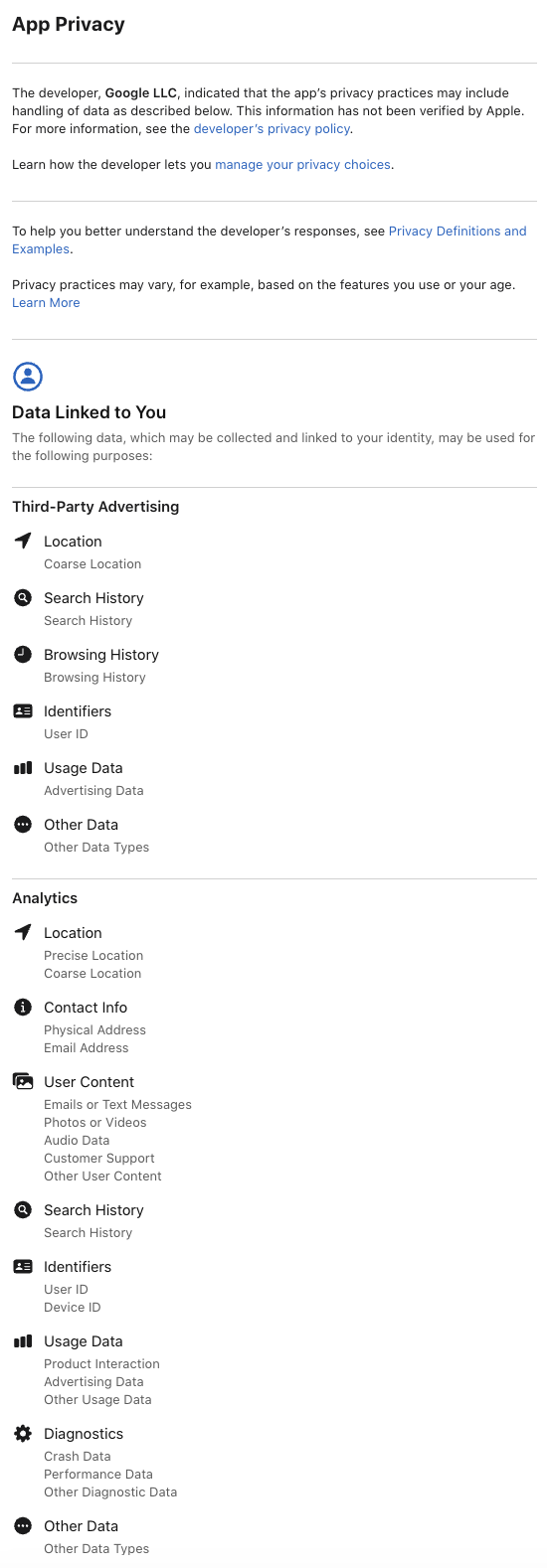}
    \end{framed}
    \caption{Expanded Privacy Label of Google Maps Part 1 (iOS)}
\end{figure}

\begin{figure}[ht!]
    \begin{framed}
        \includegraphics[width=\textwidth]{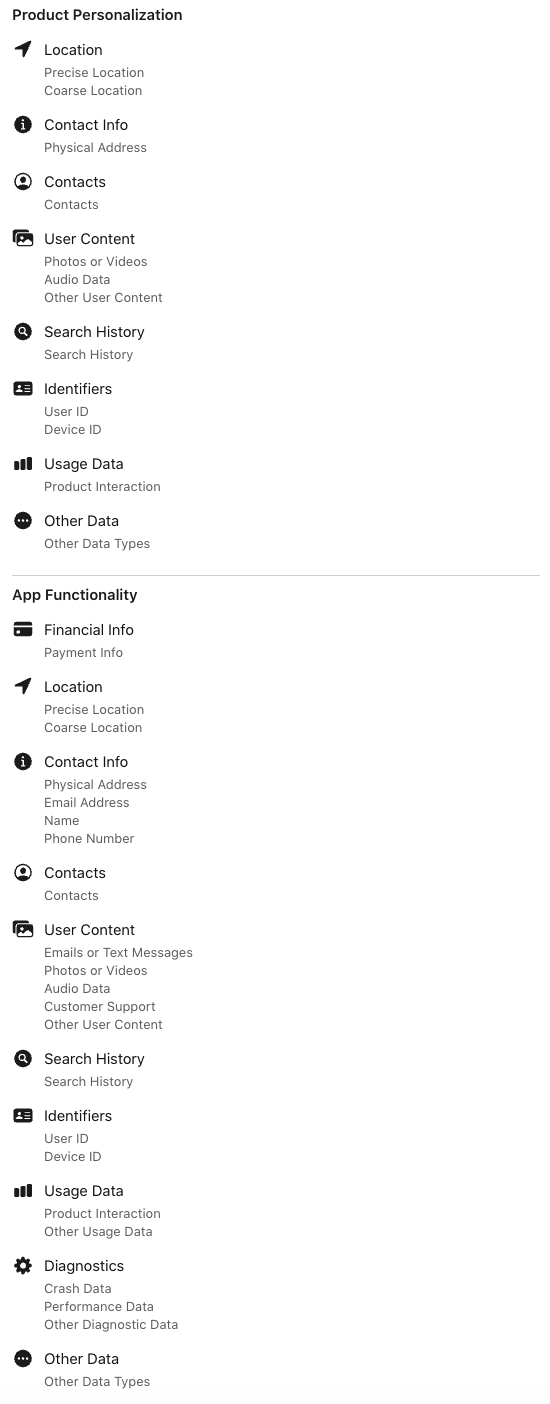}
    \end{framed}
    \caption{Expanded Privacy Label of Google Maps Part 2 (iOS)}
\end{figure}

\begin{figure}[ht!]
    \begin{framed}
        \includegraphics[width=\textwidth]{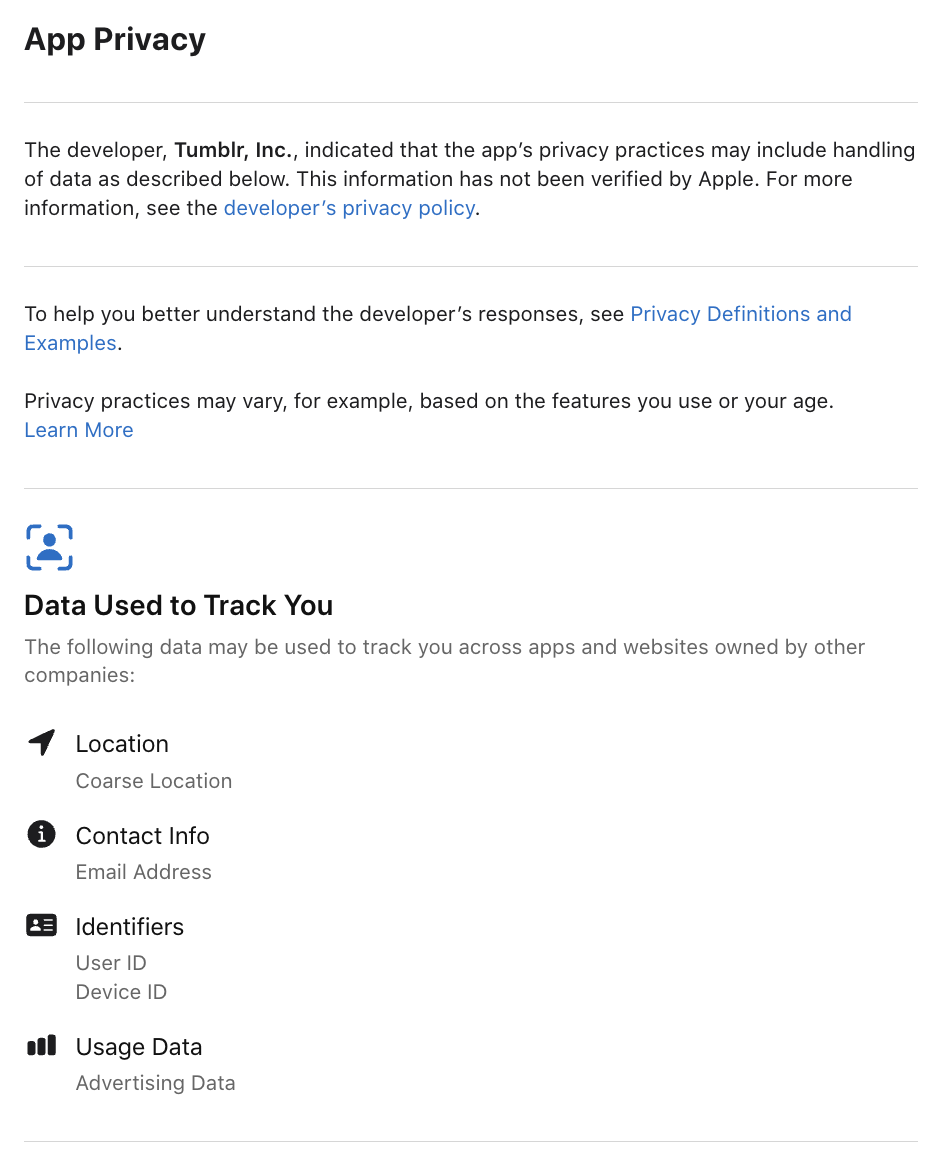}
        \includegraphics[width=\textwidth]{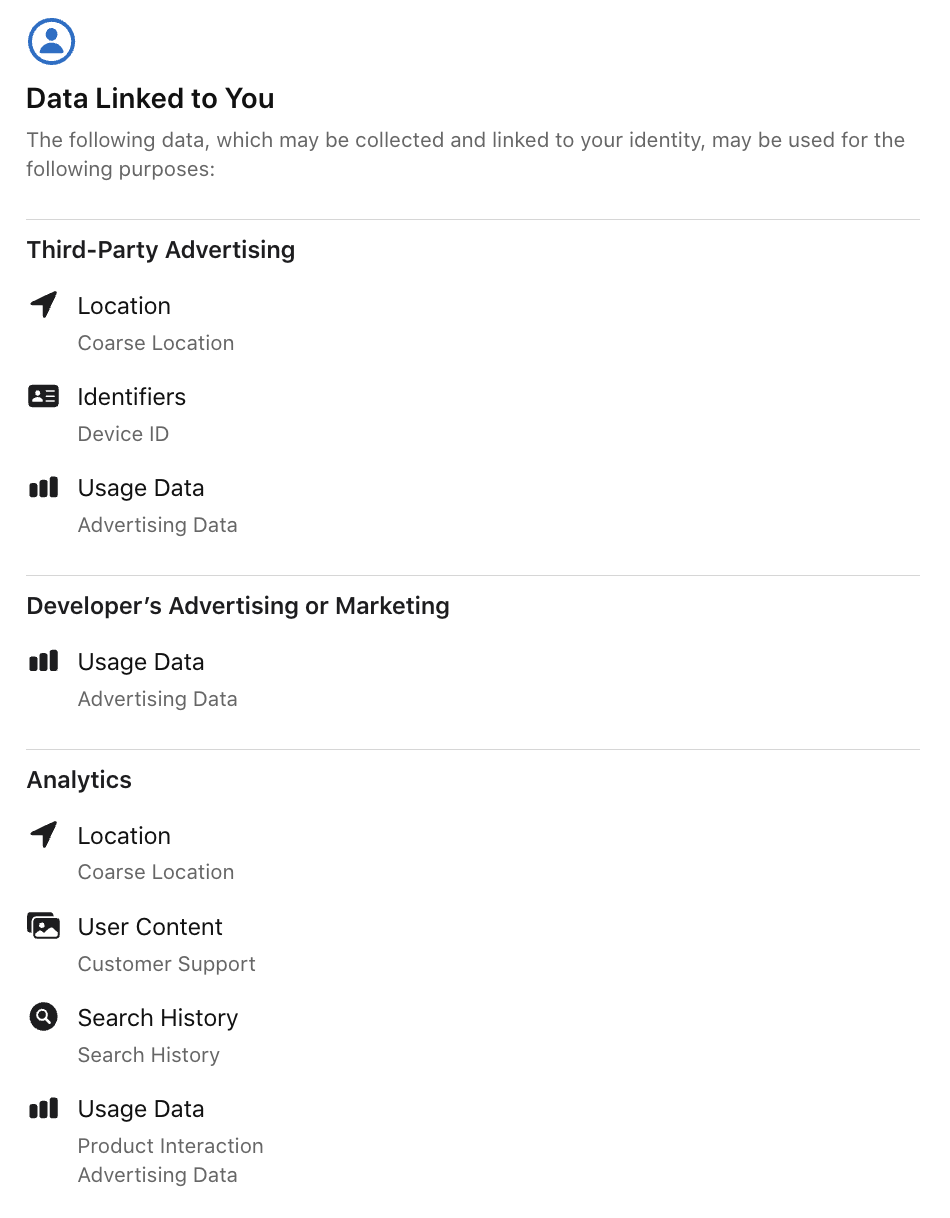}
    \end{framed}
    \caption{Expanded Privacy Label of Tumblr-Fandom, Art, Chaos Part 1 (iOS)}
\end{figure}

\begin{figure}[ht!]
    \begin{framed}
        \includegraphics[width=\textwidth]{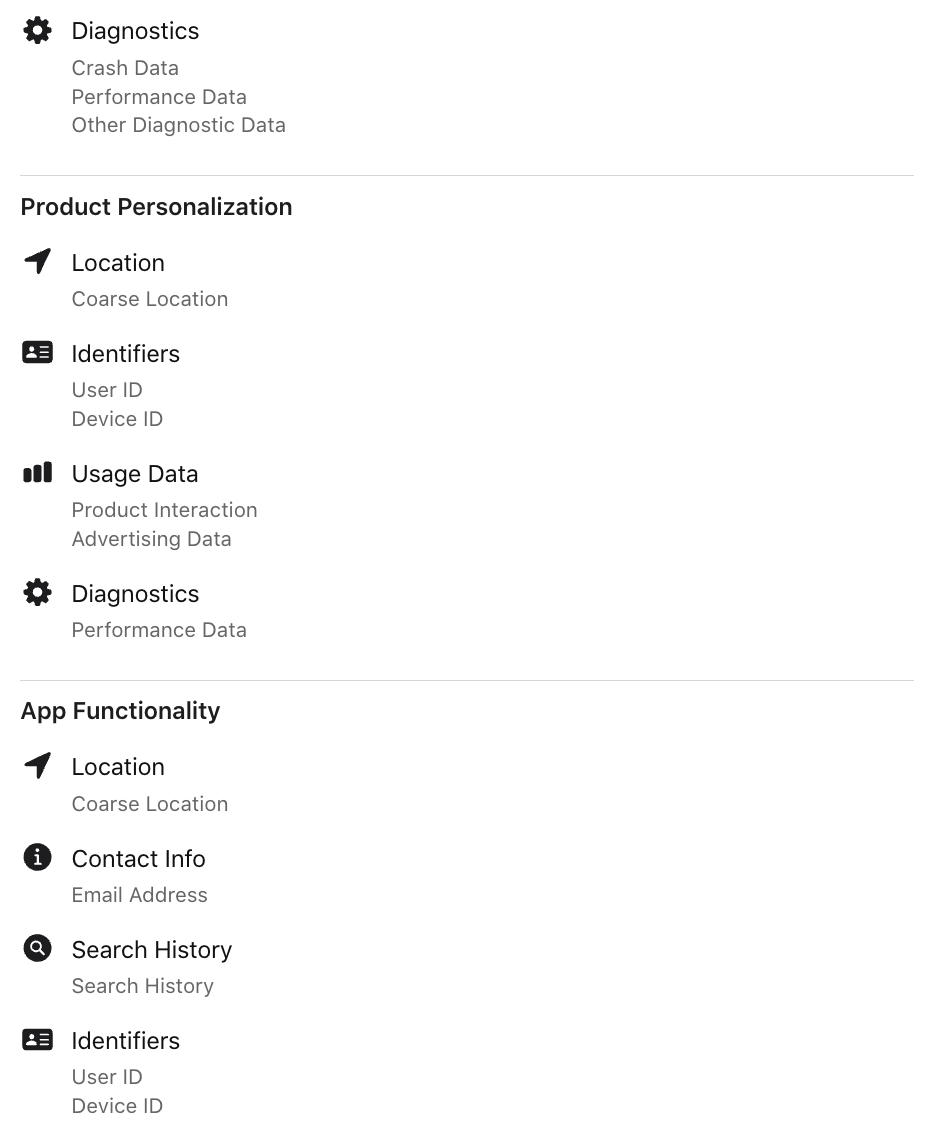}
        \includegraphics[width=\textwidth]{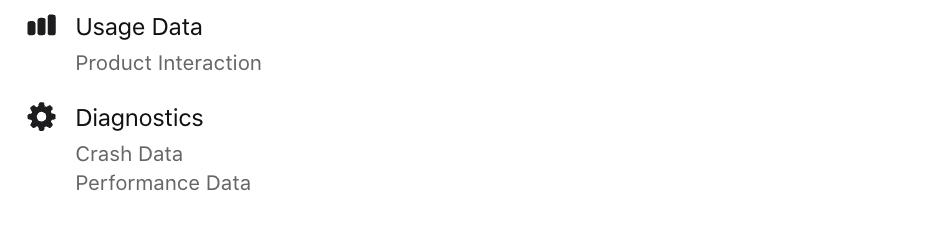}
    \end{framed}
    \caption{Expanded Privacy Label of Tumblr-Fandom, Art, Chaos Part 2 (iOS)}
\end{figure}

\end{document}